\documentclass[]{emulateapj}

\newcommand{\target}{target}
\newcommand{\control}{control}
\newcommand{\Ftarget}{\ensuremath{f_{\mathrm{\target}}}}
\newcommand{\Fcontrol}{\ensuremath{f_{\mathrm{\control}}}}
\newcommand{\Ntarget}{\ensuremath{\mathcal{N}_{\mathrm{\target}}}}
\newcommand{\Ncontrol}{\ensuremath{\mathcal{N}_{\mathrm{\control}}}}
\newcommand{\Rtarget}{\ensuremath{\mathcal{R}_{\mathrm{\target}}}}
\newcommand{\Rcontrol}{\ensuremath{\mathcal{R}_{\mathrm{\control}}}}
\newcommand{\Ha}{\ensuremath{\mathrm{H}\alpha}}
\newcommand{\composite}{composite}
\newcommand{\EBV}{\ensuremath{E(B-V)}}
\newcommand{\Q}{\ensuremath{Q}}
\newcommand{\sigv}{\ensuremath{\sigma_v}}
\newcommand{\mingals}{\ensuremath{\mathcal{N}_{\mathrm{gal}}^{\mathrm{min}}}}
\newcommand{\tauv}{\ensuremath{\tau_V}}
\newcommand{\ICM}{ICM}

\newcommand{\template}{template}
\newcommand{\average}{average}
\newcommand{\averaging}{averaging}
\newcommand{\averaged}{averaged}
\newcommand{\SNR}{SNR}

\renewcommand{\cite}[1]{\citeauthor{#1} \citeyearpar{#1}}

\begin{document}

\slugcomment{Astrophys.~J.~in press}

\title{The Transparency of Galaxy Clusters}
\shorttitle{Transparency of Galaxy Clusters}

\author{
Jo~Bovy\altaffilmark{1},
David~W.~Hogg\altaffilmark{1,2}, and
John~Moustakas\altaffilmark{1}}
\shortauthors{Bovy et al.}

\altaffiltext{1}{Center for Cosmology and Particle Physics, Department of Physics, New York 
University, 4 Washington Place, New York, NY 10003}
\altaffiltext{2}{To whom correspondence should be addressed: \texttt{david.hogg@nyu.edu}}

\begin{abstract}

If galaxy clusters contain intracluster dust, the spectra of galaxies
lying behind clusters should show attenuation by dust absorption. We
compare the optical (3500 - 7200 \AA) spectra of 60,267 luminous,
early-type galaxies selected from the Sloan Digital Sky Survey to
search for the signatures of intracluster dust in $z\sim\!0.05$ clusters. 
We select massive, quiescent (i.e., non-star-forming) galaxies using an 
EW(\Ha) $\leq 2$ \AA \ cut and consider galaxies in three bins of velocity 
dispersion, ranging from 150 to 300 km s$^{-1}$. The uniformity of early-type
galaxy spectra in the optical allows us to construct
inverse-variance-weighted \composite\ spectra with high signal-to-noise
ratio (ranging from $10^2-10^3$). We compare the \composite\ spectra of
galaxies that lie behind and adjacent to galaxy clusters and find no
convincing evidence of dust attenuation on scales $\sim\! 0.15-2$ Mpc;
we derive a generic limit of $\EBV < 3 \times 10^{-3}$ mag on
scales $\sim\! 1-2$ Mpc at the 99\% confidence level, using conservative 
jackknife error bars, corresponding to a dust mass 
$\lesssim 10^8$ $M_{\odot}$. On scales smaller than 1 Mpc this limit is 
slightly weaker, $\EBV < 8 \times 10^{-3}$ mag.

\end{abstract}

\keywords{dust, extinction --- galaxies: clusters: general --- 
intergalactic medium --- methods: statistical}

\section{Introduction}

Galaxy clusters are known to contain galaxies and hot gas, and they may 
contain extragalactic but intracluster dust, or be accreting intergalactic 
dust from their neighborhoods. Indeed, there must be some intracluster dust 
created by winds from the intracluster stars, which make up a significant 
fraction of the total stellar mass in the cluster \citep{Ferguson98}. 
Alternatively, dust could be introduced into the intracluster medium (ICM) 
through such processes as cooling-flows \citep{Fabian94}, galaxy or cluster 
mergers and collisions \citep{Popescu:2000qs}, supernovae-driven galactic 
winds \citep{Okazaki93}, ram pressure stripping of galaxies as they travel 
through the intracluster medium \citep{Gunn72} and accretion of primordial 
dust \citep{Popescu:2000qs}. Many of these processes have associated 
timescales of order 10$^8$-10$^9$ yr. A crucial question is then whether 
the dust thus injected into the intracluster medium can survive thermal 
sputtering in the hot gas. Typical dust grain sputtering timescales are 
$\tau_{\mbox{sp}} \sim\! 10^6-10^9$ yr \citep{Draine79}, similar to the 
timescales of the dust-producing processes. These timescales imply that 
only the most recently injected dust is still surviving at any given moment 
in time, from which we conclude that the amount of dust in the \ICM\ should 
be small and non-uniformly distributed.

Measurements of intracluster dust have a long and rich
history. The presence of dust was first hypothesized to explain the 
discrepancy between counts of galaxies located behind and adjacent to the 
Coma cluster \citep{Zwicky57}. A first estimate of 0.4 mag for the 
magnitude of the $B$-band extinction was suggested 
\citep[][1962]{Zwicky61}, although infrared emission of dust in the Coma
Cluster was not detected \citep{Dwek90}. Zwicky's method was improved over 
the years as catalogs of galaxy clusters became available and $B$-band 
extinctions of order 0.2 mag were reported based on various procedures:
using essentially the same approach as Zwicky, but
using a larger sample of 15 galaxy clusters \citep{Karachentsev}; 
considering color residuals to arrive at the amount of absorption
within the Local Supercluster \citep{Vaucouleurs}; looking at angular
correlations among clusters and quasars, leading to an extinction of
about 0.12 mag over radii of several Mpc \citep{Bogart}; considering
correlations of high-redshift quasars with low-redshift galaxies, which
gave evidence for dust in clusters at redshift z $\sim\! 0.15$ at a
characteristic linear radius of $500 h^{-1}$ kpc, corresponding to a
dust sphere of mass 10$^{10}$ $M_\odot$ \citep{Boyle}. However, galaxy 
number counts are subject to a variety of biases and the dearth of 
galaxies behind clusters could have other causes than 
dust \citep{Nollenberg:2003cd}.

Correlations of quasars with nearby clusters were reconsidered and a 
$B$-band extinction of 0.15 mag was found \citep{Romani:1991nd}. A similar 
result was found to explain an excess of higher redshift galaxies in nearby 
small galaxy groups \citep{Girardi1992}. However, comparing the color 
distribution of quasars behind a cluster with those in the vicinity of 
the cluster limited the relative reddening of the two samples to 
$\EBV \lesssim 0.05$ mag \citep{Maoz:1995yk}. Similar limits were 
obtained comparing color distributions of galaxies behind and removed 
from APM clusters \citep[on 1.3 Mpc scales]{Nollenberg:2003cd} and 
using large, elliptical galaxies \citep{Ferguson93}.

There have been some, but contradictory, reports on dust in the central 
regions of clusters. A study of {\it IRAS} images of 56 clusters found 
two clusters with far-infrared color excesses that could be due to 
10$^9$ $M_{\odot}$ of dust \citep{Wise93}. An average excess reddening in 
10 cooling-flow galaxy clusters of $\EBV \sim\! 0.19$ mag was reported for 
lines-of-sight to the center of these clusters \citep{Hu92}. However, a 
later report found no convincing evidence of submillimeter dust emission 
in 11 cooling-flow clusters and set an upper limit of 10$^8$ $M_\odot$ on 
the total mass of the dust \citep{Annis93}.

More recently, observations of six Abell clusters found a rough estimate of 
a dust mass of 10$^7$ $M_\odot$ in the Coma cluster, but no evidence of 
dust in the other five observed clusters \citep{Stickel:2001nt}; and no 
significant amount of infrared emission from intracluster dust in Abell 
2029 was found \citep{Bai:2007kp}. \cite{Chelouche:2007rm} reported 
reddening in a $0.1 < z < 0.3 $ sample of $\sim\! 10^4$ galaxy clusters by 
correlating the Sloan Digital Sky Survey cluster and quasar catalog and by 
comparing photometric and spectroscopic properties of quasars behind 
the clusters to those in the field. They found mean $\EBV$ values of a 
few $\times 10^{-3}$ mag for sight lines passing $\sim\!$ Mpc from the 
clusters' centers. However, a recent study found no evidence of dust in 
$0.2 < z < 0.5$ clusters from a photometric study of color excesses in 
several bands and, assuming a Galactic extinction law, derived an average 
visual extinction of $\langle A_V \rangle = 0.004 \pm 0.010$ mag 
\citep{Muller:2008hg}. 

In this paper we study the dust content of galaxy clusters by
comparing the spectra of galaxies behind clusters of galaxies with
those of galaxies not behind clusters. More specifically, we use the
optical (3500 - 7200 \AA) spectra of luminous, early-type
galaxies because they are known to dominate the stellar mass density
of the Universe \citep{Fukugita:1997bi,Hogg:2002ci} and show great
regularities in their properties \citep[e.g.][2003b, 2003c,
  2003d]{Oke68,Faber73,Visvanathan77,Djorgovski:1987vx,
  Dressler:1987ny,Kormendy:1989dg,Bower:1992jx,Roberts:1994gp,
  Bernardi03a}. Their spectra show a remarkable similarity with any
variation that does exist explained by the environment and luminosity
\citep{Eisenstein:2002ta}. Any dust-attenuation-like difference that
can be found between the composite spectra of galaxies behind galaxy
clusters and galaxies in the field can be attributed reliably to
interactions of the galaxies' light with the intracluster medium.

The difference between our study and the other precise studies cited above, 
is that (1) we have more control over our galaxy population, (2) we are
selecting on properties that only weakly involve color and are
therefore less likely to be biased, and (3) we have well-calibrated spectrophotometry
of all objects.  On the other hand, these considerations limit the size 
of our sample, so what we gain in per-object precision we lose in
number of objects, in some sense. 
Despite our relatively small sample, we obtain among the 
most stringent upper limits ever.

In what follows, AB magnitudes are used throughout, a cosmological world model with 
$(\Omega_{\mbox{m}},\Omega_\Lambda) = (0.3,0.7)$ is adopted, and the 
Hubble constant is $H_0 = 70$ km s$^{-1}$ Mpc$^{-1}$ \citep{Komatsu}, for the purposes 
of calculating distances \citep[e.g.][]{Hogg:1999ad}.

\section{Data}

The Sloan Digital Sky Survey (SDSS) is obtaining \emph{u,g,r,i} and \emph{z} 
CCD imaging of 10$^4$ deg$^2$ of the northern Galactic sky, and from that 
imaging, selecting roughly 10$^6$ targets for spectroscopy, most of them 
galaxies with $r < 17.77$ mag 
\citep[][2004, 2005]{Gunn:1998vh,York:2000gk,Stoughton:2002ae,Abazajian:2003jy}.

All the data processing, including astrometry \citep{Pier:2002iq},
source identification, deblending and photometry
\citep{Lupton:2001zb}, calibration \citep{Fukugita:1996qt,
  Smith:2002pca, Ivezic:2004bf}, spectroscopic target selection
\citep{Eisenstein:2001cq, Strauss:2002dj, Richards:2002bb},
spectroscopic fiber placement \citep{Blanton:2001yk}, spectral data
reduction and analysis (Schlegel \& Burles 2006, in preparation;
Schlegel 2006, in preparation) are performed with automated SDSS
software.

We use the spectroscopic and photometric catalog from the NYU Value Added 
Galaxy Catalog \citep[NYU-VAGC;][]{Blanton2005} compiled from the SDSS Data
Release Four \citep[DR4;][]{AdelmanMcCarthy:2005se}.

As we will be \averaging\ the spectra of galaxies coming from different 
spectral ``plates'', we depend on the calibration of these fluxes. The 
calibration procedure is as follows 
\citep[D.~J. Schlegel, in preparation;][]{Stoughton:2002ae}: Every spectral 
``plate'' of fiber positions includes several faint (15.5-18.5 mag) F8 
subdwarf stars. The spectrum of each standard star is spectrally typed by 
comparing with a grid of theoretical spectra generated from Kurucz model 
atmospheres \citep{Kurucz1992} using the spectral synthesis code SPECTRUM 
\citep{Gray1994,Gray2001}. The spectra are calibrated with these F star spectra; i.e. 
they are multiplied by the function of wavelength that makes the F star 
spectra match the F star spectrophotometry (after correcting for Galactic reddening). 
This calibration procedure produces consistent calibration at the 
5\% level\protect\footnote{$\sim\!4.4$\% in $g$-$r$ and $\sim\!2.8$\% in $r$-$i$ for galaxies, see 
\url{http://www.sdss.org/dr4/products/spectra/spectrophotometry.html}}.
In addition to this, the SDSS does not use an atmospheric refraction corrector, 
so the effective fiber position on the sky shifts slightly as a function of 
wavelength. In the presence of brightness gradients, this creates a fluxing 
error.

Redshifts are measured on the reduced spectra by an automated system, which 
models each galaxy spectrum as a linear combination of stellar eigenspectra 
(D.~J. Schlegel, in preparation). The central velocity dispersion \sigv\ is 
determined by fitting the detailed spectral shape as a velocity-smoothed sum 
of stellar spectra (D.~J. Schlegel \& D.~P. Finkbeiner, in preparation).

The measurements of the equivalent width (EW) of the \Ha\ line is measured 
exactly as described in \cite{Quintero:2003we}. Briefly, a linear fit of the 
spectral section to a linear combination of the mean SDSS old galaxy spectrum 
and the mean SDSS A-star spectrum with the locations of possible emission 
lines marked out is performed; this best fit model is then scaled down to 
have the same flux continuum as the data in the vicinity of the \Ha\ emission 
line and subtracted to leave a continuum-subtracted line spectrum; the \Ha\ 
line flux is then measured in a 20 \AA \ width interval centered on the line 
and converted to a rest-frame equivalent width with a continuum found by 
taking the inverse-variance-weighted average of two sections of the spectrum 
about 150 \AA \ in size and on either side of the emission line. This method 
fairly accurately models the absorption trough in the continuum, although in 
detail it leaves small negative residuals.

The galaxy clusters used here, are $0.015 < z < 0.067$ member clusters
taken from a friends-of-friends cluster catalog constructed from the
SDSS DR3 main sample galaxies with absolute magnitudes $M_{0.1r}$ $<$
-19.9 mag \citep{Berlind:2006bx}. We first consider a \mingals = 10
member minimum, corresponding to a total absolute r-band magnitude
M$_r \leq -21.9$; for reference, the Virgo cluster contains $13$
galaxies brighter than this limit \citep{Trentham:2002mq}.  However,
we also vary the minimum number of galaxies in the cluster between
\mingals = 5 (M$_r \leq -20.8$ mag) and \mingals = 20 (M$_r \leq -23.1$ mag).

\section{Analysis}

\subsection{Sample Construction}

Our technique is analogous to the classic ``foreground screen'' test
of using stars to measure the Galactic extinction curve
\citep[e.g.,][and references therein]{Calzetti01}.  Here, we compare
the spectra of galaxies that lie behind a cluster of galaxies,
i.e. galaxies whose light has had to traverse a galaxy cluster on its
way to observational astronomers, against the spectra of galaxies that
do not lie behind a cluster. Galaxies that lie behind a galaxy
cluster will constitute the \emph{\target}\ galaxies and galaxies that
are not behind any cluster will form the \emph{\control}\ sample.  The
optical spectra of these objects must be intrinsically similar so that
we can ascribe any measured differences to dust attenuation as it
passes through a cluster of galaxies; massive, early-type galaxies are
known to have exactly this property \citep{Eisenstein:2002ta}.

\begin{figure}%
\begin{center}
\includegraphics[width=0.25\textwidth,clip=]{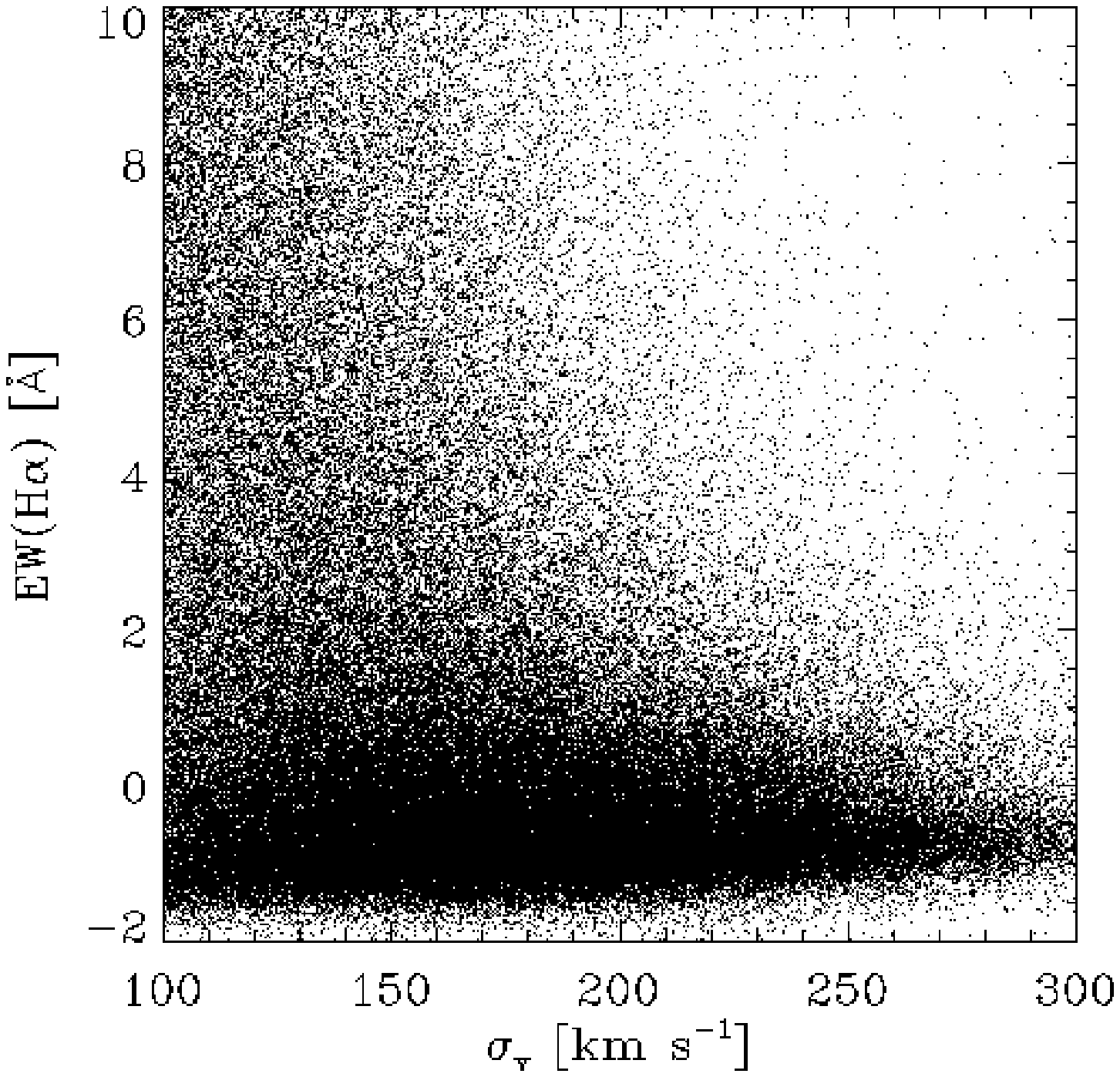}%
\includegraphics[width=0.25\textwidth,clip=]{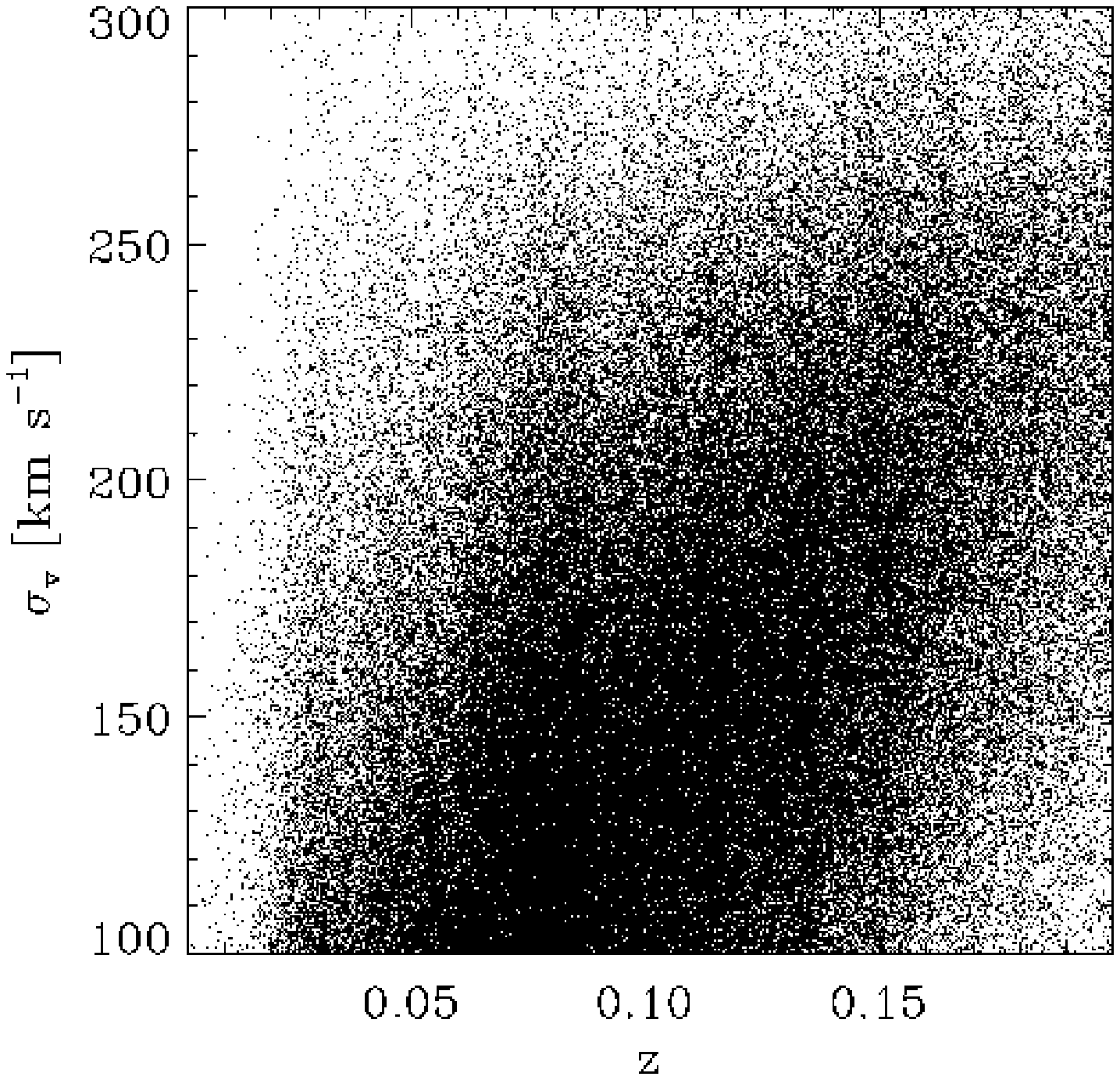}\\
\includegraphics[width=0.25\textwidth,clip=]{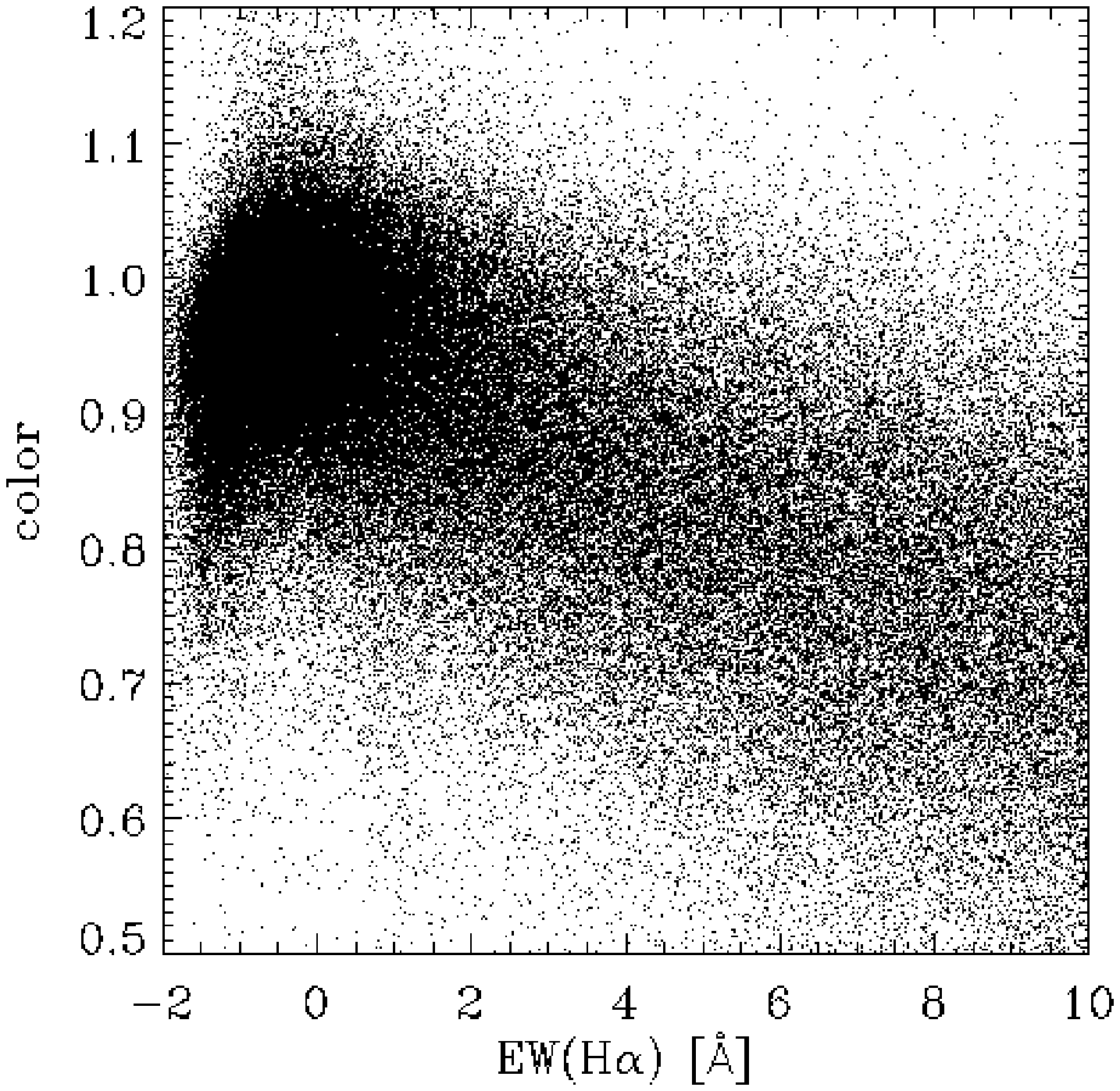}%
\includegraphics[width=0.25\textwidth,clip=]{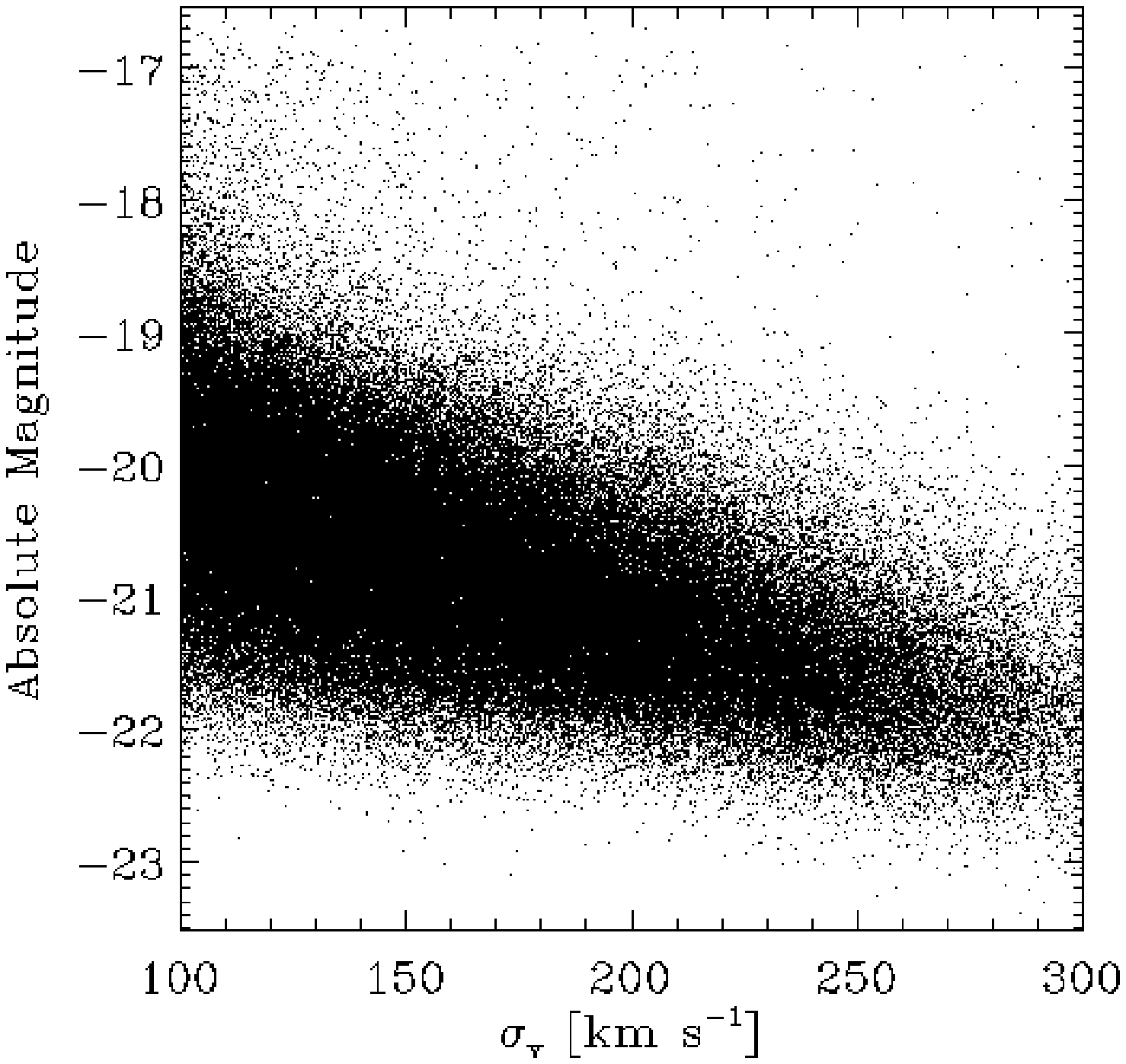}%
\caption{Properties of galaxies in the SDSS sample: Plots of the quantities relevant for the sample selection. From left to right, top to bottom: equivalent width of the \Ha\ line (EW(\Ha)) vs. velocity dispersion \sigv; \sigv\ vs. redshift; $g$-$r$ color vs. EW(\Ha); $r$-band absolute magnitude vs. \sigv.}%
\label{galprop}
\end{center}
\end{figure}

\begin{deluxetable*}{cccccc}
\tablecaption{Properties of the various samples\label{tablesamples}}
\tablecolumns{6}
\tablewidth{0pt}
\tablehead{
\colhead{Sample$^a$} & \colhead{EW(\Ha)} & \colhead{\sigv} & \colhead{\mingals$\,^b$} & \colhead{\Rtarget$^c$} & \colhead{\Rcontrol$^d$}\\
\colhead{} & \colhead{(\AA)} & \colhead{(km s$^{-1}$)} & \colhead{} & \colhead{(Mpc)} & \colhead{(Mpc)}}
\startdata
  Primary & $\leq$ 2 & $200 \leq \sigv \leq 250$ & 10 & 0.50 & 2 \\
  $150 \leq \sigv \leq 200$ & $\leq$ 2 & $150 \leq \sigv \leq 200$ & 10 & 0.50 & 2\\
  $250 \leq \sigv \leq 300$ & $\leq$ 2 & $250 \leq \sigv \leq 300$ & 10 & 0.50 & 2\\
  \mingals = 5 & $\leq$ 2 & $200 \leq \sigv \leq 250$ & 5 & 0.50 & 2\\
  \mingals = 20 & $\leq$ 2 & $200 \leq \sigv \leq 250$ & 20 & 0.50 & 2\\
  \Rtarget = 0.25 Mpc & $\leq$ 2 & $200 \leq \sigv \leq 250$ & 10 & 0.25 & 2\\ 
  \Rtarget = 1 Mpc & $\leq$ 2 & $200 \leq \sigv \leq 250$ & 10 & 1.00 & 2
\enddata
\tablenotetext{a}{The redshift range for all these samples is $0.1  \leq z \leq 0.2$.}
\tablenotetext{b}{Minimum number of members to define a galaxy cluster.}
\tablenotetext{c}{Galaxies within a transverse distance \Rtarget\ of a cluster (and are at a larger redshift) are considered behind that cluster and make up the target subsample.}
\tablenotetext{d}{Galaxies more than \Rcontrol\ from every cluster make up the control subsample.}
\end{deluxetable*}

Luminous, early-type galaxies are part of the red sequence of galaxies
and can be identified in many different ways. Here, we select galaxies
from the SDSS sample based on two properties derived from their
optical spectra: the equivalent width of the \Ha\ emission line,
EW(\Ha), and the stellar velocity dispersion, \sigv. We select
quiescent (i.e., non-star-forming) galaxies using EW(\Ha)$<2$~\AA.  In
addition, we restrict the sample to the redshift range $0.1 \leq z
\leq 0.2$. Figure \ref{galprop} shows plots of EW(\Ha), \sigv, color,
and absolute magnitude of relevant galaxies in the SDSS sample.

The galaxy clusters used to define our \target\ and
\control\ subsamples were taken from a friends-of-friends cluster
catalog \citep{Berlind:2006bx}. We emphasize that this catalog is not
complete and that we only use part of the catalog to select our
\target\ and \control\ galaxies. In addition to this, our galaxy spectra 
come from SDSS DR4, which has a larger coverage than SDSS DR3, out of which 
the cluster catalog was constructed. We expect the number of ``false
negatives'', i.e. galaxies that might be catalogued as
\control\ galaxies that are actually behind a cluster, to be small.
In addition, because of the large number of galaxies in the
\control\ sample, the effect of any contaminating or misclassified
galaxies will be diluted in the stacking procedure.  Finally, we will
see that the principal source of errors will be due to the
\target\ galaxies, for which this effect is not important.

We can now specify when a galaxy is part of the \target\ subsample of
galaxies and when it belongs to the \control\ subsample. For each
galaxy cluster, every galaxy at a redshift exceeding that of the
cluster that is found to be within 0.5 Mpc transverse distance of it,
is considered to be behind the cluster. We define \Rtarget = 0.5 Mpc,
which we will vary to investigate the radial dependence of any
measured effect. Galaxies that are more than \Rcontrol\ = 2 Mpc
removed from every galaxy cluster, are classified as
\control\ galaxies. Any galaxy between these bounds is excluded from 
the analysis.

\begin{figure}%
\centering
\includegraphics[width=0.25\textwidth]{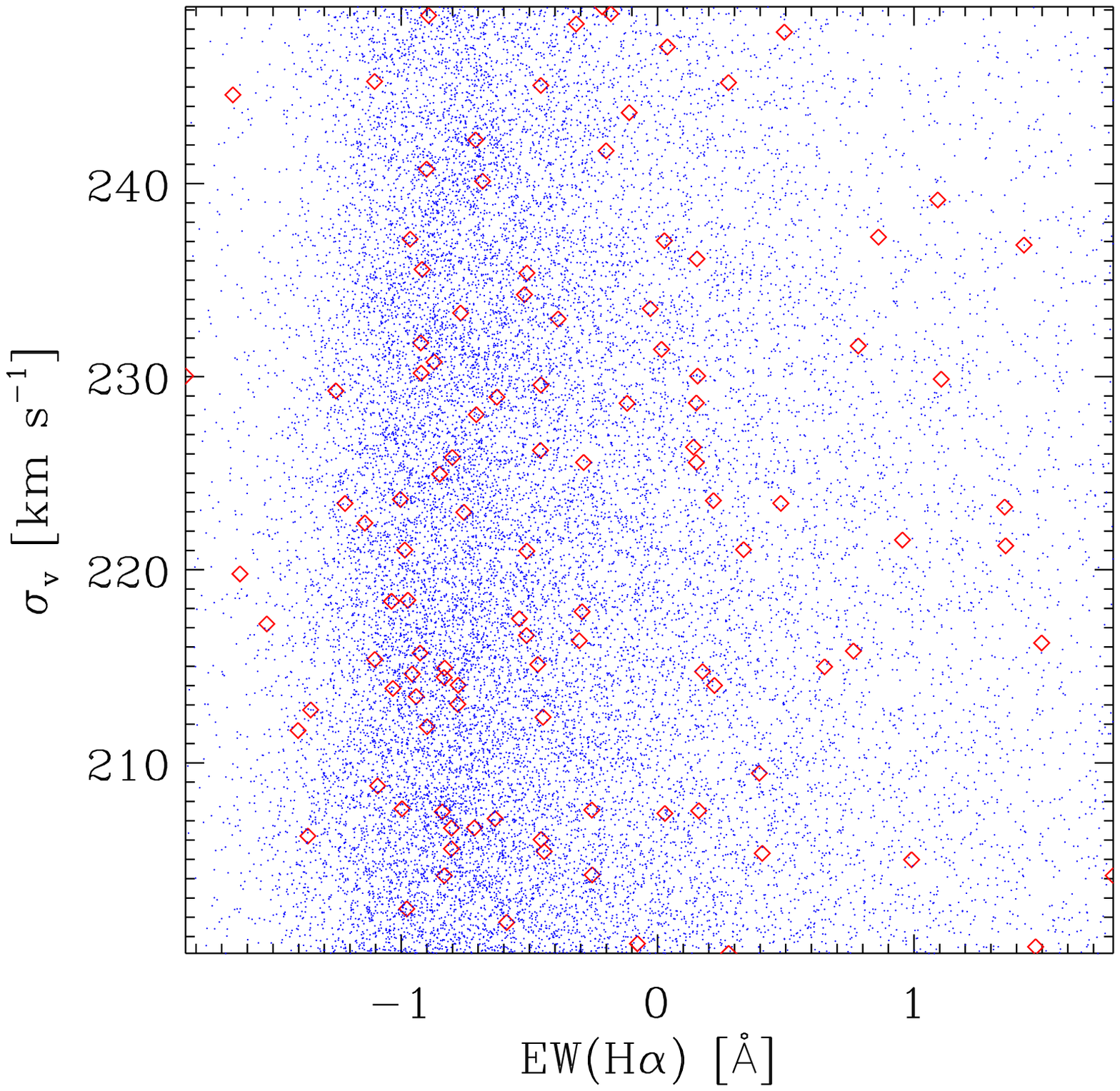}%
\includegraphics[width=0.25\textwidth]{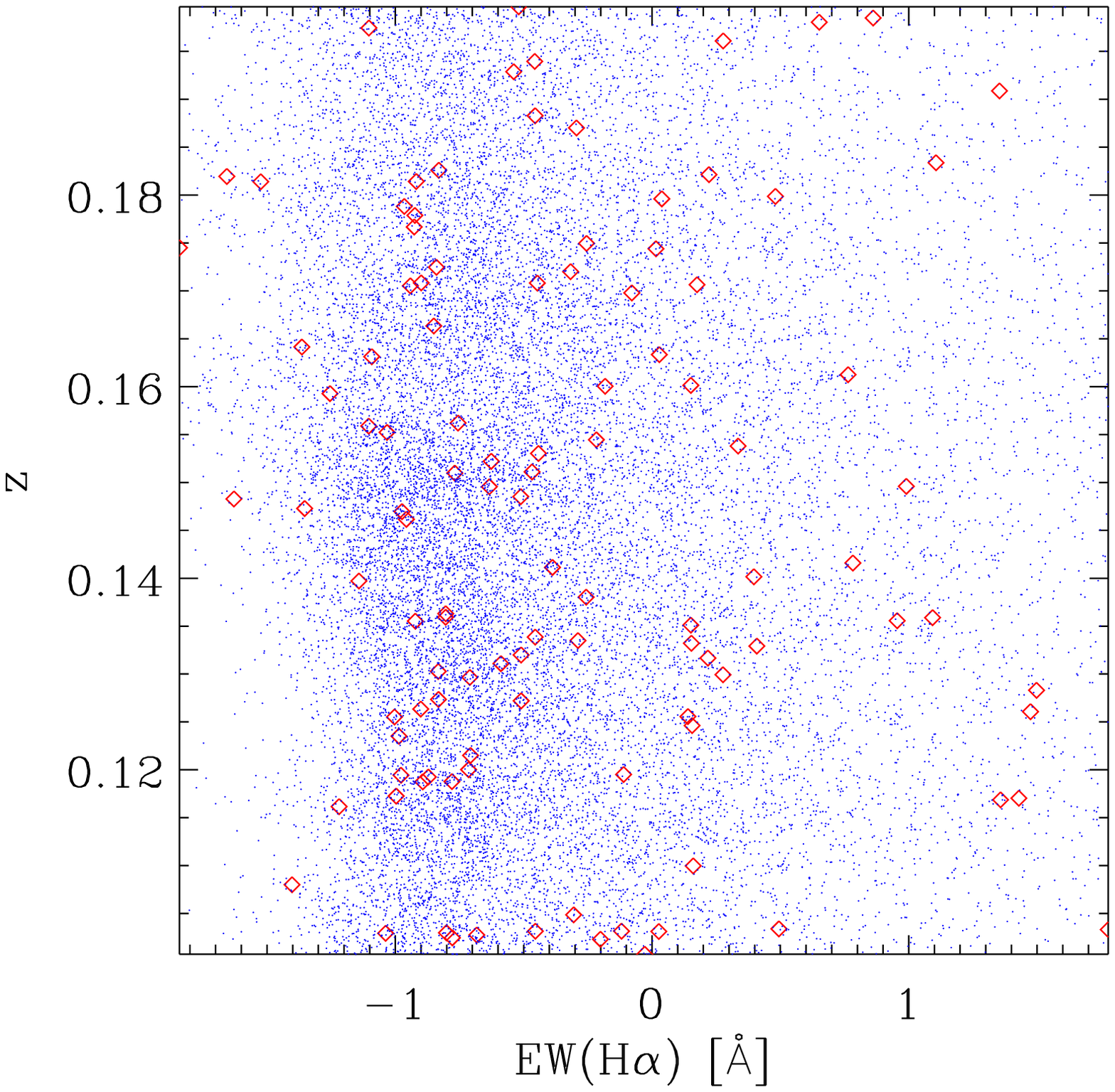}\\
\includegraphics[width=0.25\textwidth]{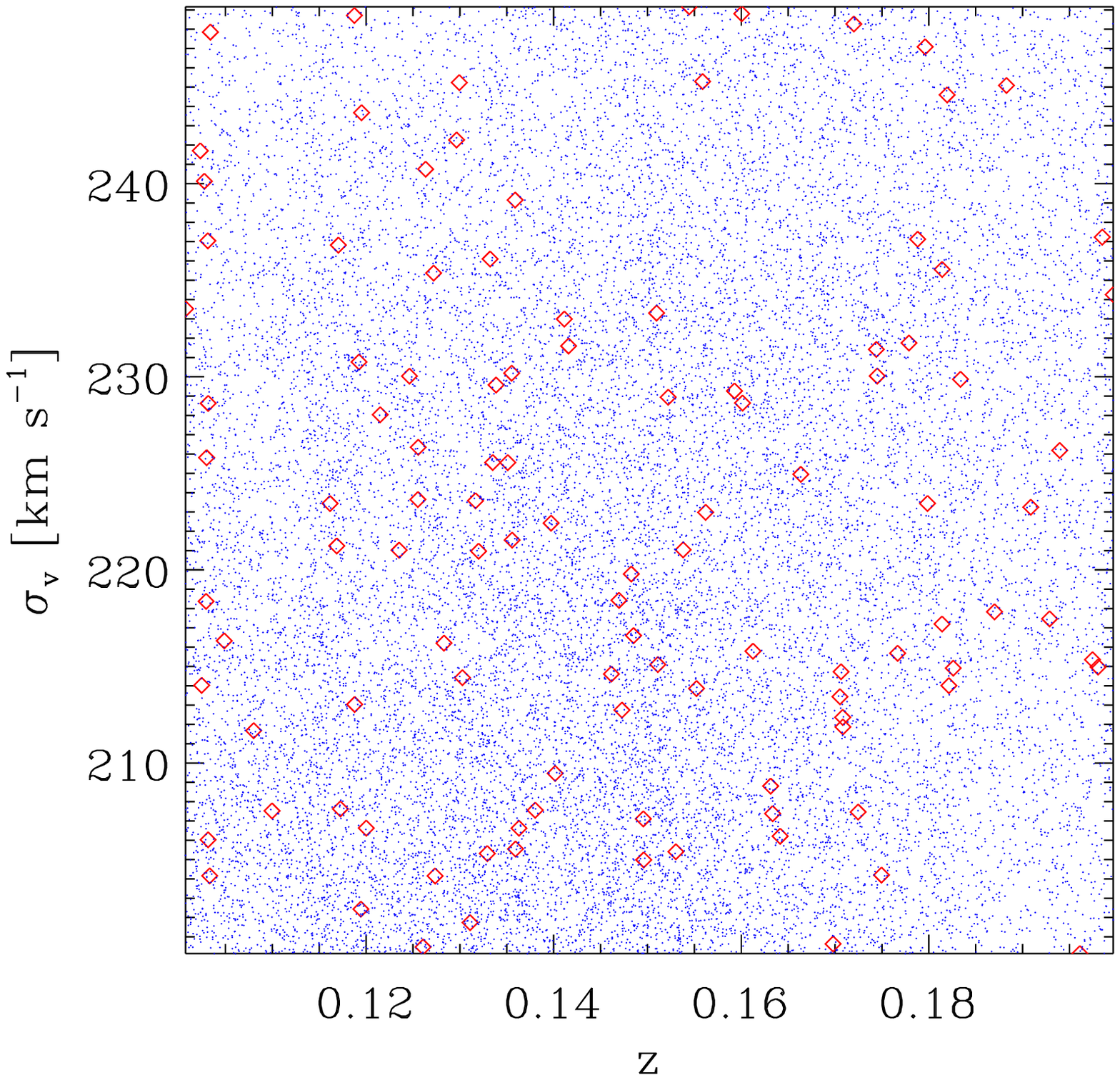}%
\includegraphics[width=0.25\textwidth]{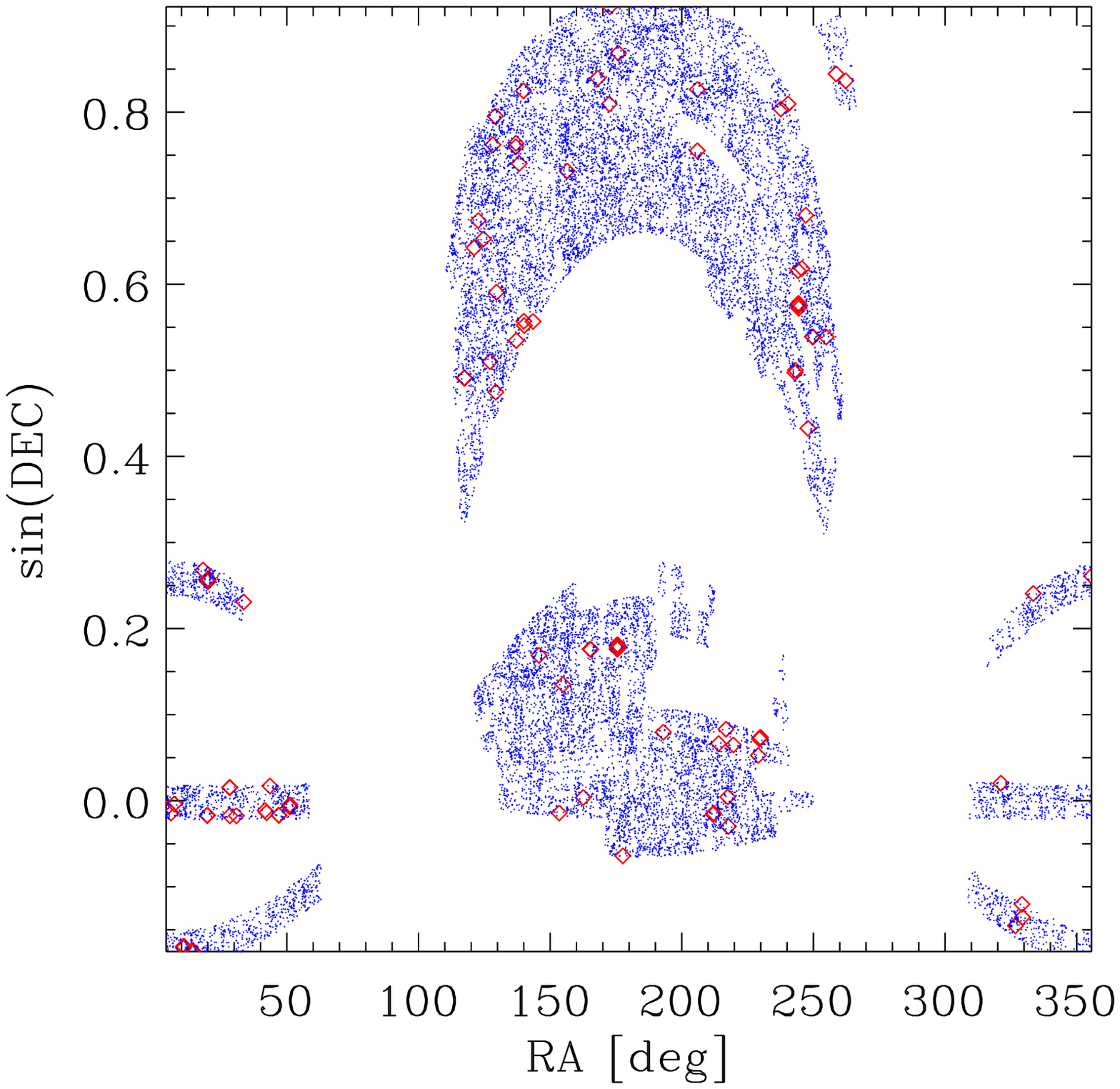}%
\caption{Galaxy properties for the primary sample: Plots of the quantities relevant for the sample selection for the primary sample (see Table \ref{tablesamples} for the definition of the primary sample). The \target\ subsample is represented by diamonds $\diamond$, the \control\ subsample by dots. From left to right, top to bottom: velocity dispersion vs. EW(\Ha); redshift vs. EW(\Ha); velocity dispersion vs. redshift; angular distribution of the \target\ and \control\ subsamples: RA vs. sin(DEC).}%
\label{mainprop}
\end{figure}

Based on these criteria a \emph{primary} sample with \target\ and
\control\ subsamples was constructed as follows: galaxies with EW(\Ha)
$\leq 2$ \AA, $200 \leq \sigv \leq 250$ km s$^{-1}$ and $0.1 \leq z
\leq 0.2$ were classified as \target\ or \control\ based on the values
\Rtarget\ = 0.5 Mpc and \Rcontrol\ = 2 Mpc, using groups from the
cluster catalog with a minimum of 10 members. Figure \ref{mainprop}
shows plots of several of these properties for the primary sample as
well as their distribution on the celestial sphere. This figure
illustrates that the distribution of \target\ and \control\ galaxies
is approximately uniform in these properties. Several secondary
samples were also considered, based on variations of some of these
properties, i.e. \sigv, \mingals\ and \Rtarget\ (see
Table~\ref{tablesamples} for an overview of the properties of the
various samples used). 

Added together, the total number of galaxy spectra
that we consider in this paper is 60,267. This number includes the number 
of galaxies in the primary and secondary samples, broken down in 
Table~\ref{tableresults} by subsample (note, however, that there is a 
significant overlap of used spectra between certain samples), as well as the 
sizes of the \target\ subsamples used in the determination of the radial 
dependency of dust-attenuation (see below).

\newpage
\subsection{Stacking Procedure and Comparison}

Since we are creating \composite\ spectra from individual spectra that 
are intrinsically very similar, we will weight the pixels of each 
spectrum with their inverse variance, as this will 
give us the highest signal-to-noise \composite\ spectrum. However, 
the overall normalization of each spectrum is determined by its 
distance from us and therefore the spectra are only alike when properly 
normalized. This can be achieved by comparing each spectrum to a common 
\template. This comparison consists 
of a simple, one-parameter fit to determine the scale factor between 
the \template\ and the spectrum to be added, which is used to rescale 
the spectrum before adding it to the weighted \average.

\begin{figure*}%
\centering
\includegraphics[width=0.3\textwidth]{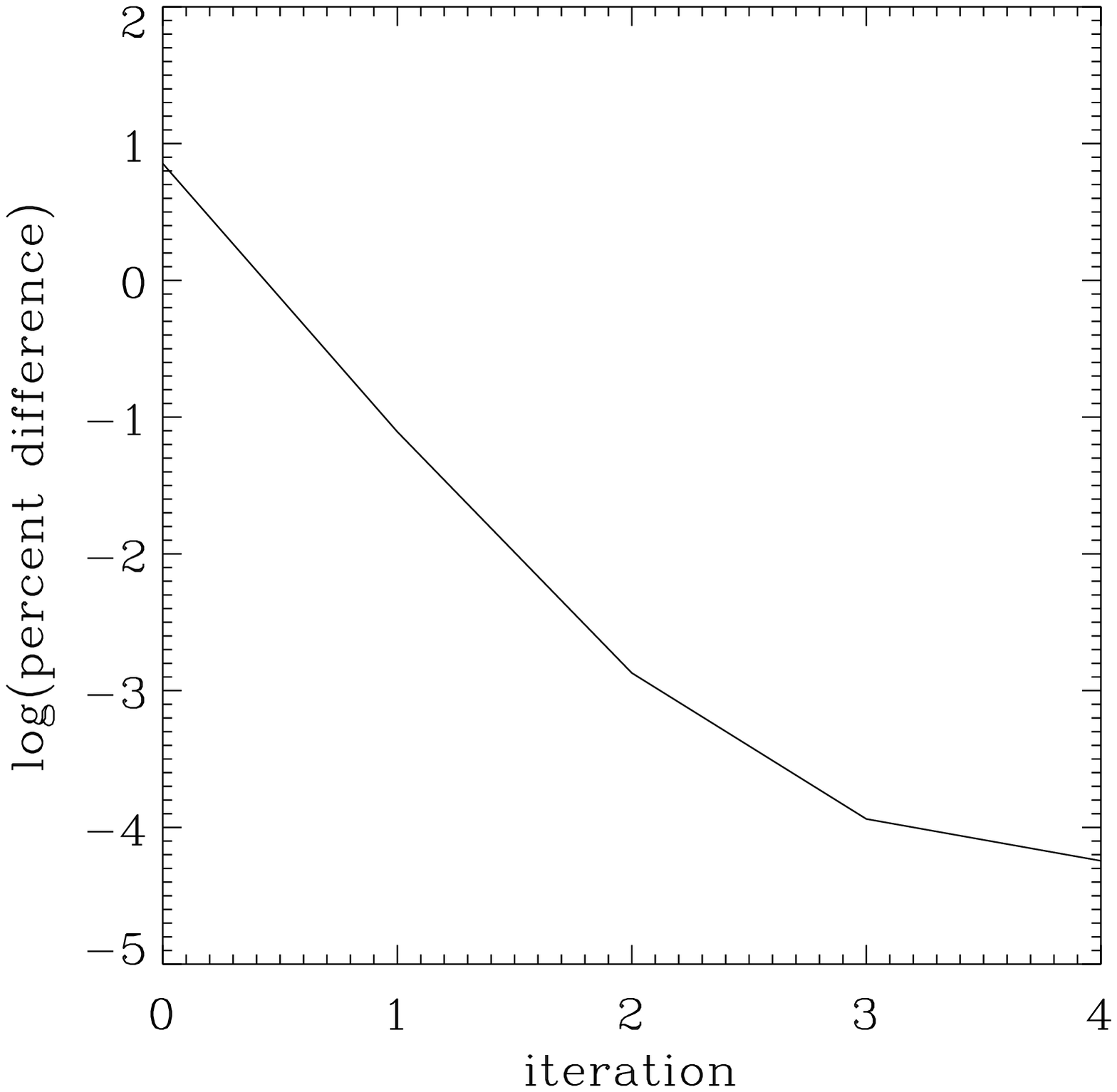}%
\includegraphics[width=0.3\textwidth]{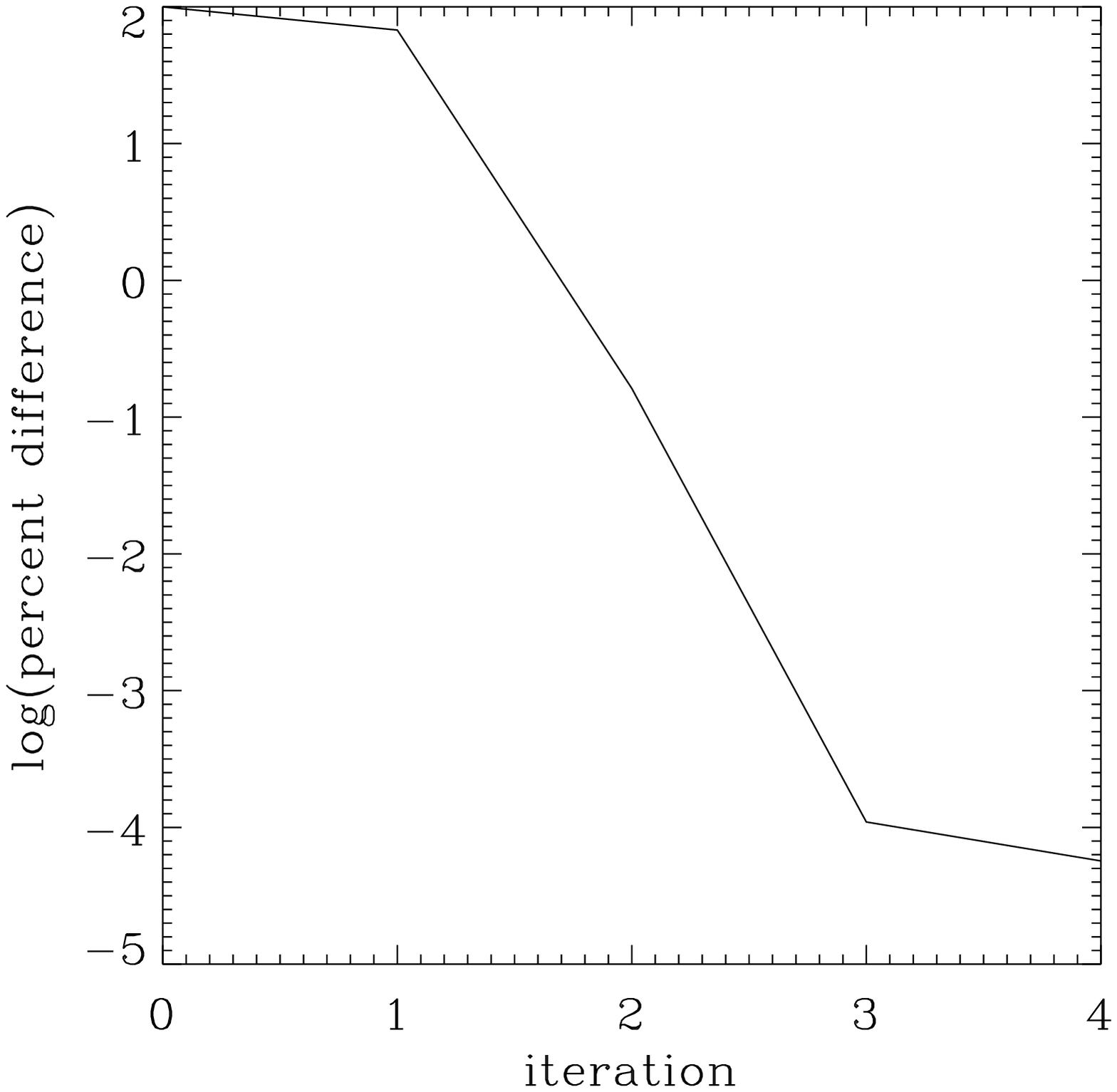}%
\includegraphics[width=0.3\textwidth]{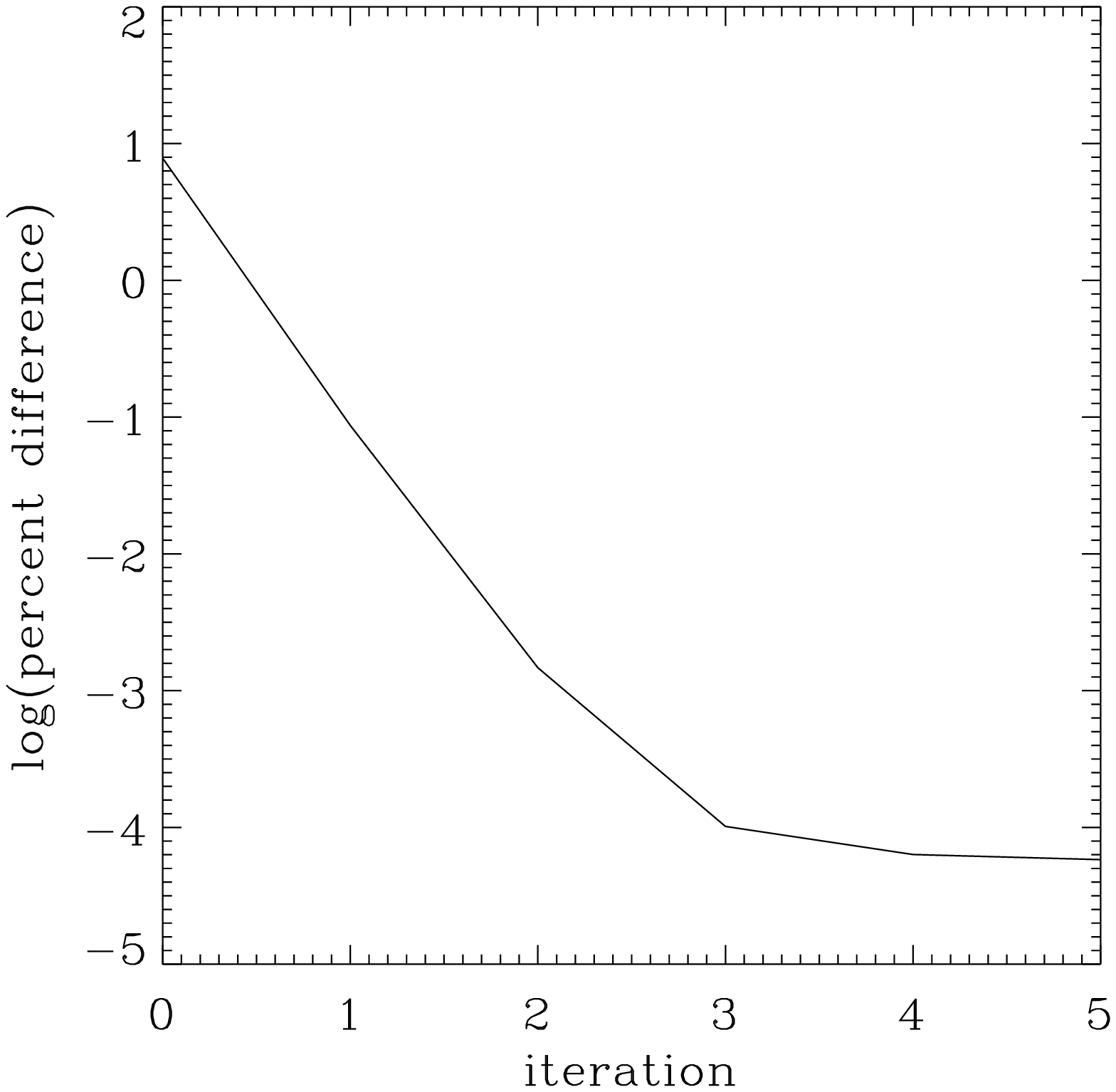}%
\caption{Convergence of the \average\ of the \control\ spectra for the different velocity dispersion bins: From left to right: $200 \leq \sigv \leq 250$, $150 \leq \sigv \leq 200$, and $250 \leq \sigv \leq 300$.}%
\label{mainconvergence}
\end{figure*}

When stacking the spectra in the \control\ subsample, such a \template\ 
is not available a priori. Therefore, we have designed a procedure to 
self-consistently create this \template\ from the constituent spectra, 
using a simple, iterative algorithm. To start off, the spectra are 
averaged as given, using inverse-variance-weighting, with no individual rescaling 
applied. In the next step, this first \average\ features as the \template\ 
which is used to scale each spectrum before adding it. The resultant \average\ 
becomes the new \template\ in the next iteration. This continues until 
the difference between successive iterations, defined as the maximum 
fractional difference over the spectrum, is less than $\sim\!1$ part in 
10$^{-6}$. Convergence is achieved after about 5 iterations and can be 
quite dramatic (see Fig.~\ref{mainconvergence}, which shows this 
convergence for the primary sample and the other velocity 
dispersion bins). The upshot of this iterative procedure is that 
the resulting \average\ is essentially the same as the \template\ used to 
compute it, i.e.~the \composite\ \control\ spectrum is obtained using itself 
as template. Finally, the \composite\ spectrum is 
smoothed using a Gaussian filter with a velocity width equal to the 
maximum velocity dispersion of the sample.

We use this \composite\ \control\ spectrum as the template when 
obtaining the \composite\ \target\ spectrum, since the spectra in these 
subsamples are intrinsically very similar. Indeed, the \composite\ 
\control\ spectrum is the best \template\ to use in the sum of the \target\ 
spectra, as it is a very high signal-to-noise representation of each 
\target\ spectrum, and setting the overall normalization of each \target\ 
spectrum equal to that of the \composite\ \control\ spectrum will ensure that 
small wavelength-dependent, attenuation-like differences will be the primary 
difference between the \composite\ \target\ spectrum and the \composite\ 
\control\ spectrum, i.e.~the overall normalization of the \composite\ 
spectrum of the \target\ spectra will be very close to that of the \composite\ 
\control\ spectrum. The \composite\ \target\ spectrum obtained in this way 
is smoothed using the same Gaussian filter as used on the \composite\ 
\control\ spectrum.

Finally, we fit to

\begin{equation}\label{eqcomp}
\Ftarget(\lambda) = \Q \Fcontrol(\lambda)  \ \mathrm{e}^{-\tau(\lambda)} , 
\end{equation}

\noindent where \Ftarget\ and \Fcontrol\ is the \composite\ spectrum of the
\target\ and \control\ subsamples, respectively, and \Q\ is an arbitrary
scale factor. Following \cite{Charlot00}, we parameterize the dust
attenuation law as 

\begin{equation}
\tau(\lambda) = \tauv \Big(\frac{\lambda}{5500 \mbox{\AA}} \Big)^{-\alpha} \ ,
\end{equation}

\noindent where \tauv\ is the $V$-band optical depth.  As in
\cite{Charlot00}, we adopt $\alpha = 0.7$, which is a reasonable
approximation to the shape of the Milky Way optical extinction curve.
Note that our parameterization ignores the possibility that the
emission of the cluster itself is influencing our results; however,
since we expect the cluster light to be similar to the light of early-type
galaxies, this should not bias our conclusions.

\subsection{Error Estimation}

The error on the individual pixels of the \average\ spectra 
follows immediately from the stacking procedure. The 
inverse variance of an inverse-variance-weighted \average\ is given by 
the sum of the individual weights. The \composite\ 
\target\ spectra obtained have a median signal-to-noise ratio (\SNR) of 
$\sim\!200$, while the larger \control\ subsamples lead to 
\composite\ spectra with a median \SNR\ of $\sim\!2500$. Therefore, it is 
clear that the main source of error is due to the \composite\ \target\ 
spectrum.

The error $\sigma_{\tauv}$ on the value of \tauv\ is obtained by a jackknife 
procedure. In general, a jackknife estimate of the variance of a statistic is 
obtained by dividing the sample into a number of (equal size) subsamples and 
obtaining the relevant statistic for each of these subsamples. The estimate 
of the variance of the statistic is then approximately equal to the variance 
of the values obtained for the subsamples \citep[with a proportionality constant 
that depends on the number of subsamples, which rapidly converges to unity as 
the number of subsamples increases; ][]{Efron}. Theoretically, this estimate is obtained 
from subsamples created by leaving out one of the ``datapoints'' (in our 
case, a ``datapoint'' is the spectrum of a galaxy, consisting of many 
individual points); however, due to computational constraints, this 
calculation is not always feasible. A possible way of dealing with this 
limitation is by dividing the sample into a number of subsets, based on a 
property that is unrelated to the relevant statistic, and creating jackknife 
subsamples as unions of all but one of these subsets \citep{Shao}.

Whenever a full-fledged jackknife estimate was deemed
too computationally intensive, an equal number of quantiles in
declination were chosen to subdivide the sample, and a minimum of 200
jackknife subsamples was used to calculate errors in all these cases. Since
the error is mostly due to inaccuracies in the \composite\ \target\ spectrum, 
jackknife subsamples were only created using the \target\ subsample; i.e. 
for the purpose of the jackknife procedure, the \composite\ \control\ spectrum 
was supposed to be known exactly.

As a check on the validity of our error estimates, we performed a careful 
examination of the error estimates for the primary sample. Jackknife error 
estimates were obtained using different numbers of subsamples, ranging from 
ten to the maximum, 110, the number of galaxies in the 
primary \target\ subsample, see below. Additionally, we implemented a 
bootstrap procedure, which works much in the same way as the jackknife 
procedure, but creates bootstrap samples from randomly picking 
``datapoints'' with replacement from the set of, in this case, \target\ spectra. Values up 
to 300 for the number of bootstrap samples were used. All jackknife and 
bootstrap estimates agreed on the first two significant figures of the error.

\subsection{Algorithm tests}\label{secmock}

In order to examine the consistency of the stacking procedure and the 
estimation of the error on \tauv, we have designed two algorithm-tests, 
which also tell us about the precision with which we can perform our measurement. 
First, we tested the accuracy and precision with which our stacking algorithm 
could recover a known value of the reddening, \tauv. We selected a random 
subset of 200 objects from the \control\ subsample of the primary sample 
and reddened them with a value of \tauv\ = 0.025 with Gaussian noise of 
standard deviation 0.027 (i.~e., mimicking the result for the primary sample, 
see below). We found that our stacking algorithm retrieved a value 
of \tauv\ = 0.024, with a 1$\sigma$-error of 0.038. Similar results were found 
for different input-reddening values, however, the errors computed using the 
jackknife procedure were consistently larger, but of the same magnitude as the 
variation that went in.

To better simulate the actual parameter estimation for the primary sample, 
a larger random sample of 8000 spectra from the \control\ subsample was chosen 
to be a mock-\control-subsample and a randomly chosen subsample of this 
of one hundred spectra was artificially reddened to provide the 
mock-\target-subsample. Several orders of magnitude of reddening were 
tried and we found that for values of \tauv\ of $\sim\!10^{-1}$ and 
Gaussian noise of the same magnitude, our algorithm returns the exact 
amount of reddening and error, which means that our algorithm should be 
able to detect reddening of this magnitude in the real samples. 
Values of \tauv\ of order 10$^{-2}$ were recovered by our algorithm as 
well, however, the jackknife estimate of the error is consistently larger 
than the variation that was put in. We were unable to recover values of 
\tauv\ of order 10$^{-3}$ and noise of the same magnitude. The computed 
error in this case is still of the order 10$^{-2}$, which indicates that 
this is a lower bound set by measurement errors. Any errors obtained that 
are larger than this, are likely due to an intrinsic variation in dust 
absorption among clusters, while errors of this size and smaller are 
consistent with being true measurement errors.

\section{Results}

\begin{figure}
\centering
\plotone{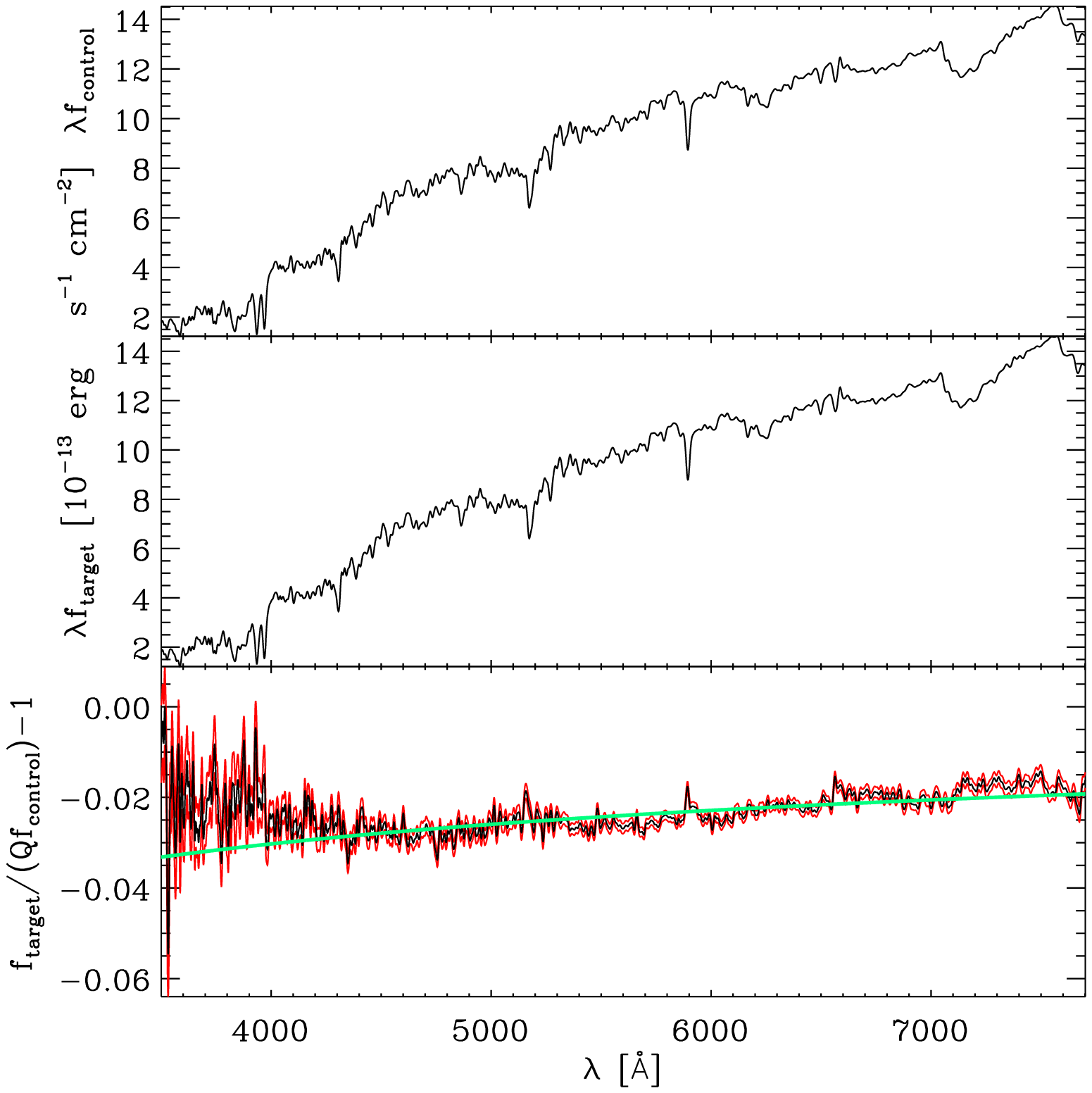}
\caption{Result for the primary sample: \average\ and comparison for the subsamples of the primary sample: the top panel shows the inverse-variance-weighted \average\ spectrum of the galaxies in the \control\ subsample, multiplied by the wavelength; the middle panel shows the same for the galaxies in the \target\ subsample; the bottom panel shows a comparison of these two \composite\ spectra (which should be exactly equal to zero if there were no dust absorption in galaxy clusters and the \averaged\ spectra were exact), and a fit to a standard dust law (see the text for more details on this procedure). The bottom panel shows error bars (the top curve is an upper bound and the bottom curve a lower bound). Errors on the \composite\ spectra in both the top panel and the middle panel are of the order of the line thickness.}%
\label{mainresult}
\end{figure}

\begin{deluxetable}{ccccccc}[b]
\tablecaption{Main results for the various samples\label{tableresults}}
\tablecolumns{5}
\tablewidth{0pt}
\tablehead{\colhead{Sample} & \colhead{\Ntarget} & \colhead{\Ncontrol} & \colhead{\tauv} & \colhead{$\sigma_{\tauv}$}}
\startdata
  Primary & 110 & 21468 & 0.025 & 0.027\\
  $150 \leq \sigv \leq 200$ & 122 & 26531 & 0.045 & 0.029\\
  $250 \leq \sigv \leq 300$ & 32 & 7239 & -0.047 & 0.053\\
  \mingals = 5 & 425 & 17993 & 0.027 & 0.014\\
  \mingals = 20 & 29 & 22562 & -0.012 & 0.059\\
  \Rtarget = 0.25 Mpc & 31 & 21468 & 0.037 & 0.039\\
  \Rtarget = 1 Mpc & 392 & 21468 & -0.001 & 0.013
\enddata
\tablecomments{Results of the comparison of the \average\ of the \target\ galaxies and the \average\ of the \control\ galaxies for the primary sample and the secondary samples. \Ntarget\ gives the number of galaxies in the \target\ subsample; \Ncontrol\ gives the number of galaxies in the \control\ subsample; \tauv\ is the best fit parameter to the dust law (eq. \ref{eqcomp}); 1$\sigma$-errors on \tauv\ are given in the $\sigma_{\tauv}$ column.}
\end{deluxetable}

Figure \ref{mainresult} shows the result of \averaging\ the spectra in the 
\control\ subsample (top panel) and of \averaging\ the spectra in the 
\target\ subsample (middle panel) for the primary sample. The lower 
panel shows a difference plot of these two quantities, with the fractional difference 
defined as

\begin{equation}
\frac{\Ftarget - \Q \Fcontrol}{\Q \Fcontrol} \ ,
\end{equation}

\noindent which for the dust attenuation in equation (\ref{eqcomp}) 
equals e$^{-\tau(\lambda)} - 1$. A fit to equation (\ref{eqcomp}) gives 
the value of \tauv, which together with the sizes of the samples 
is given in Table \ref{tableresults}. Only the calculated error on the 
difference in the lower panel is shown here. This error is substantial and 
the error in \tauv\ is, likewise, not negligible (see Table 
\ref{tableresults} for the error on \tauv).

\begin{figure*}%
\centering
\includegraphics[height=0.3\textwidth]{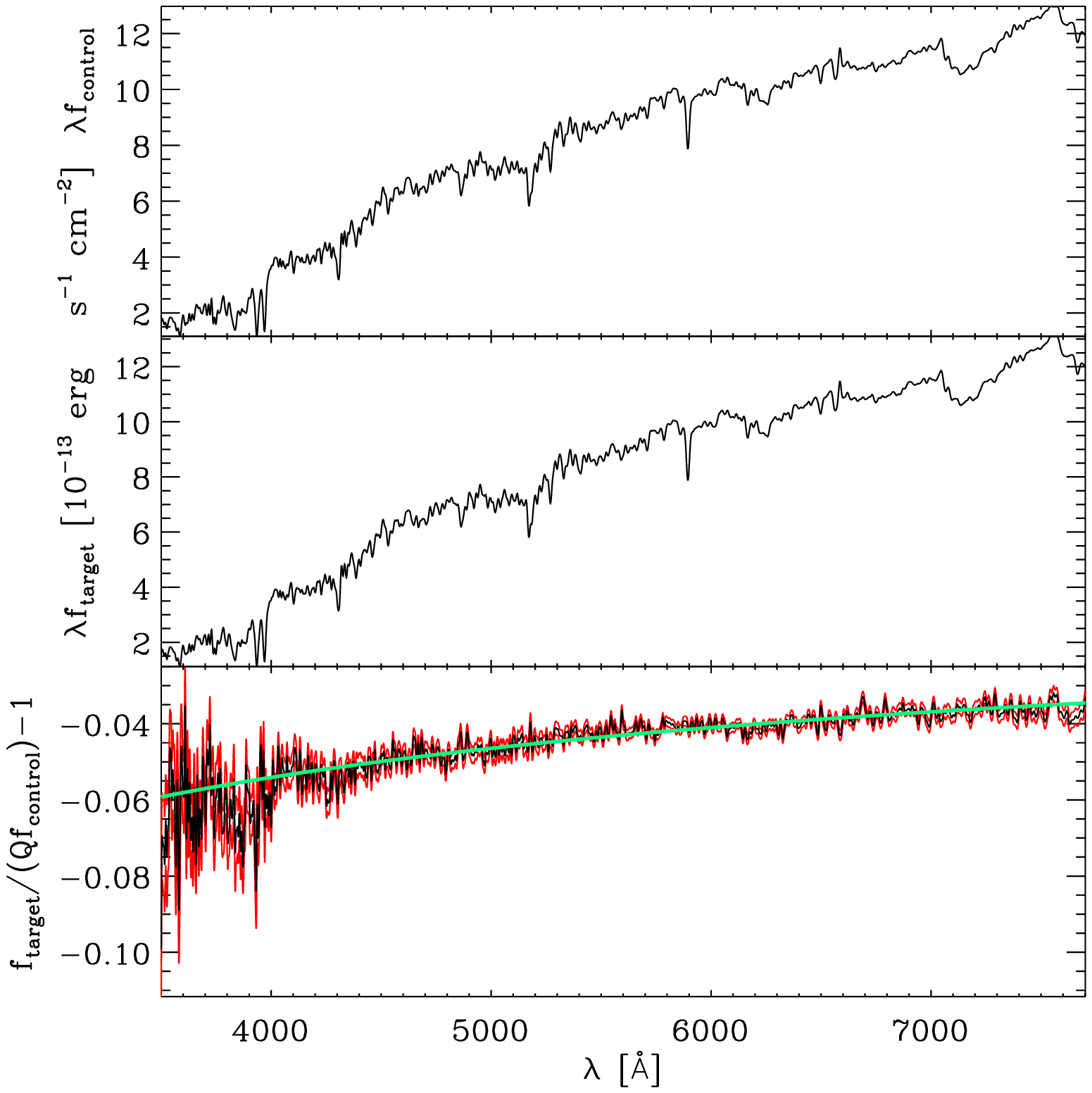}%
\includegraphics[height=0.3\textwidth]{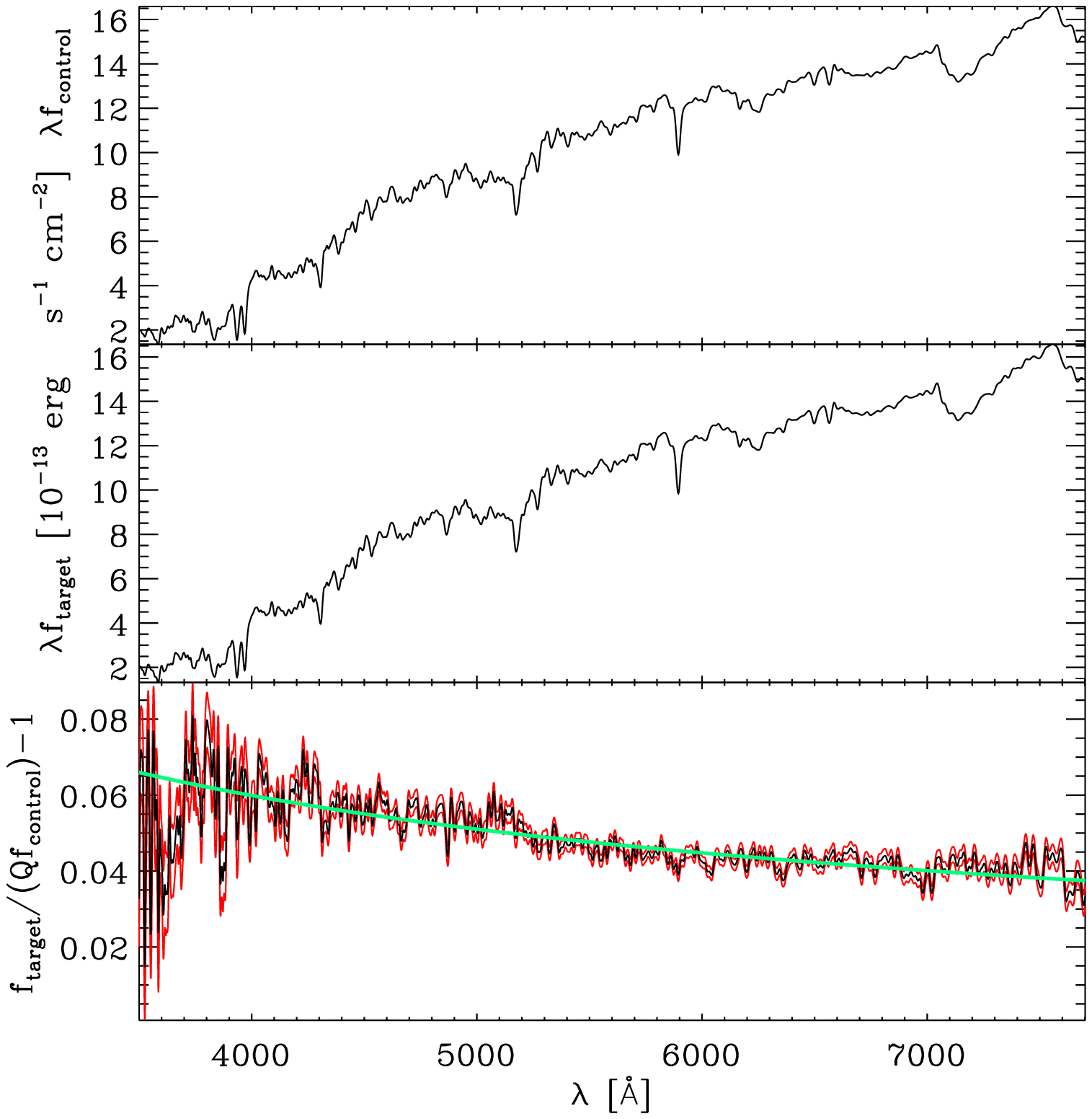}%
\includegraphics[height=0.3\textwidth]{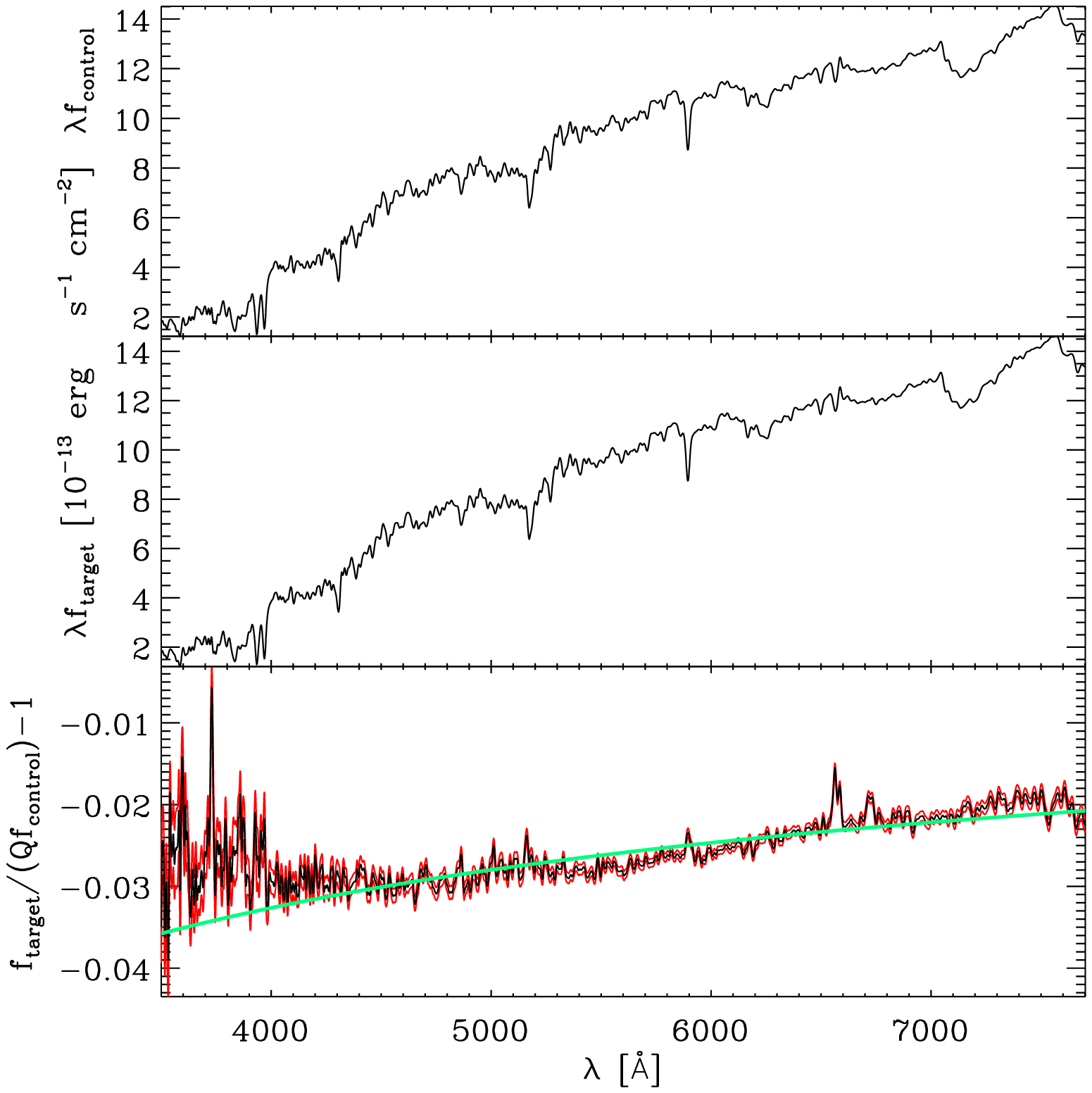}\\
\includegraphics[height=0.3\textwidth]{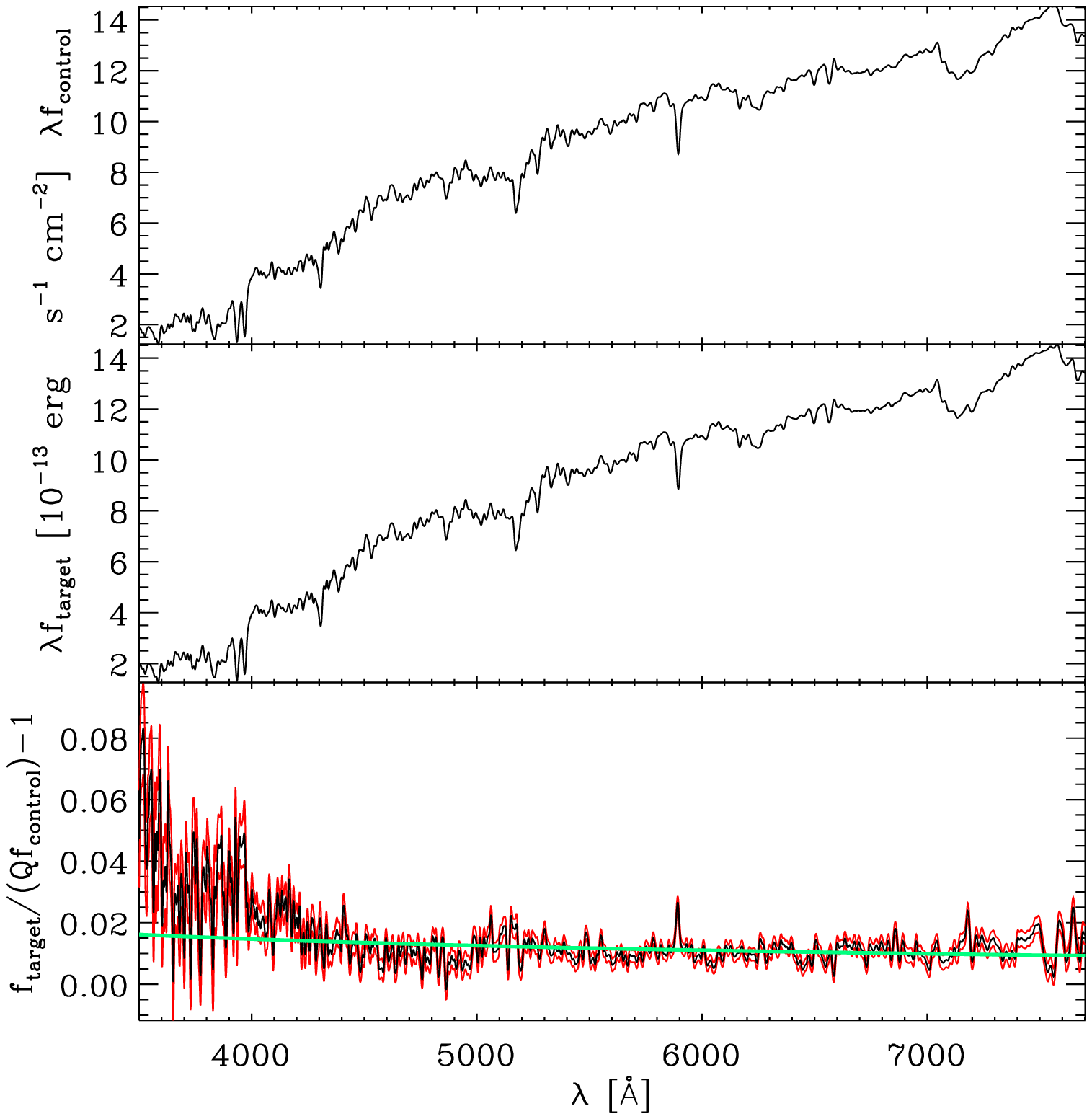}%
\includegraphics[height=0.3\textwidth]{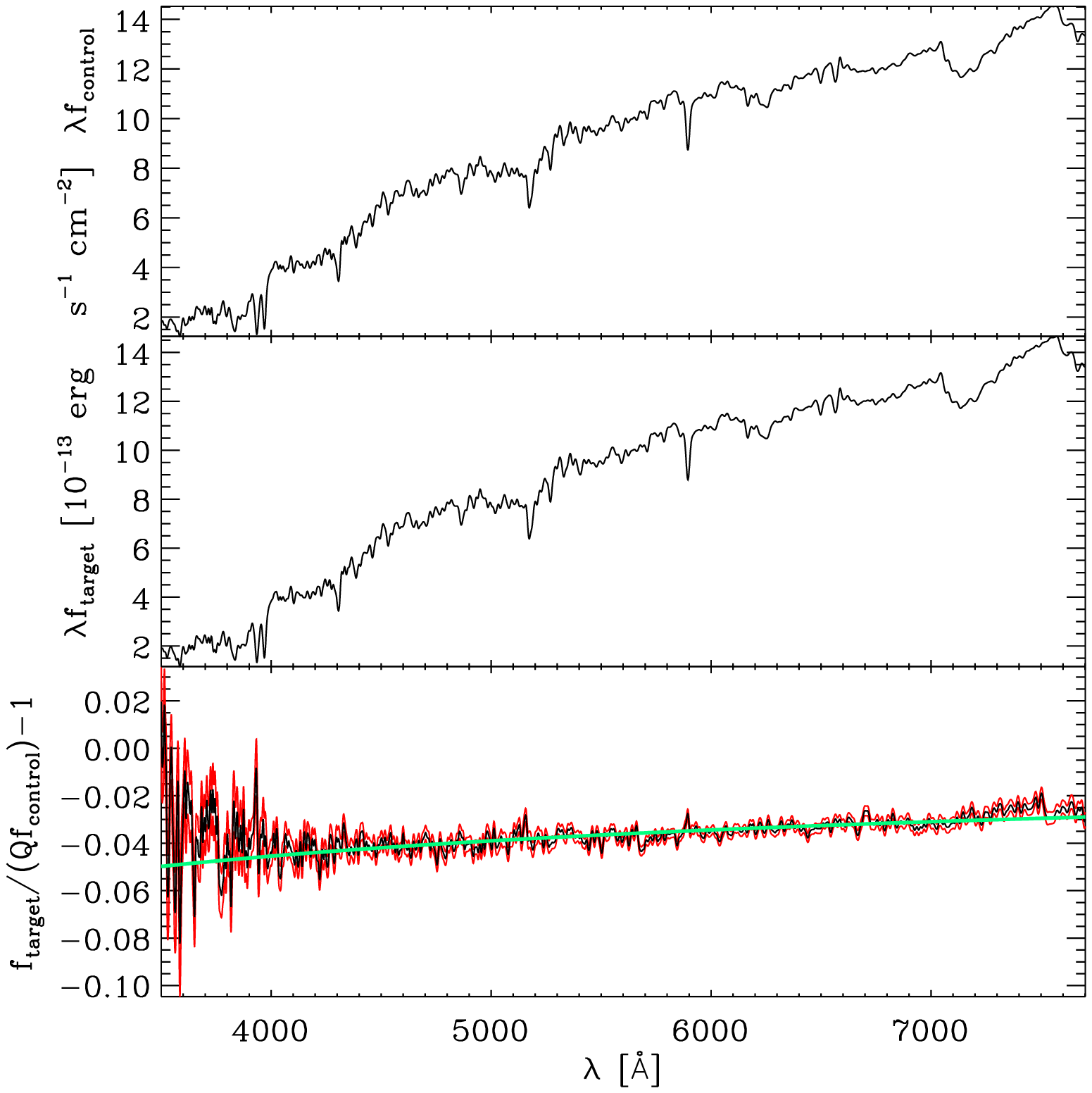}%
\includegraphics[height=0.3\textwidth]{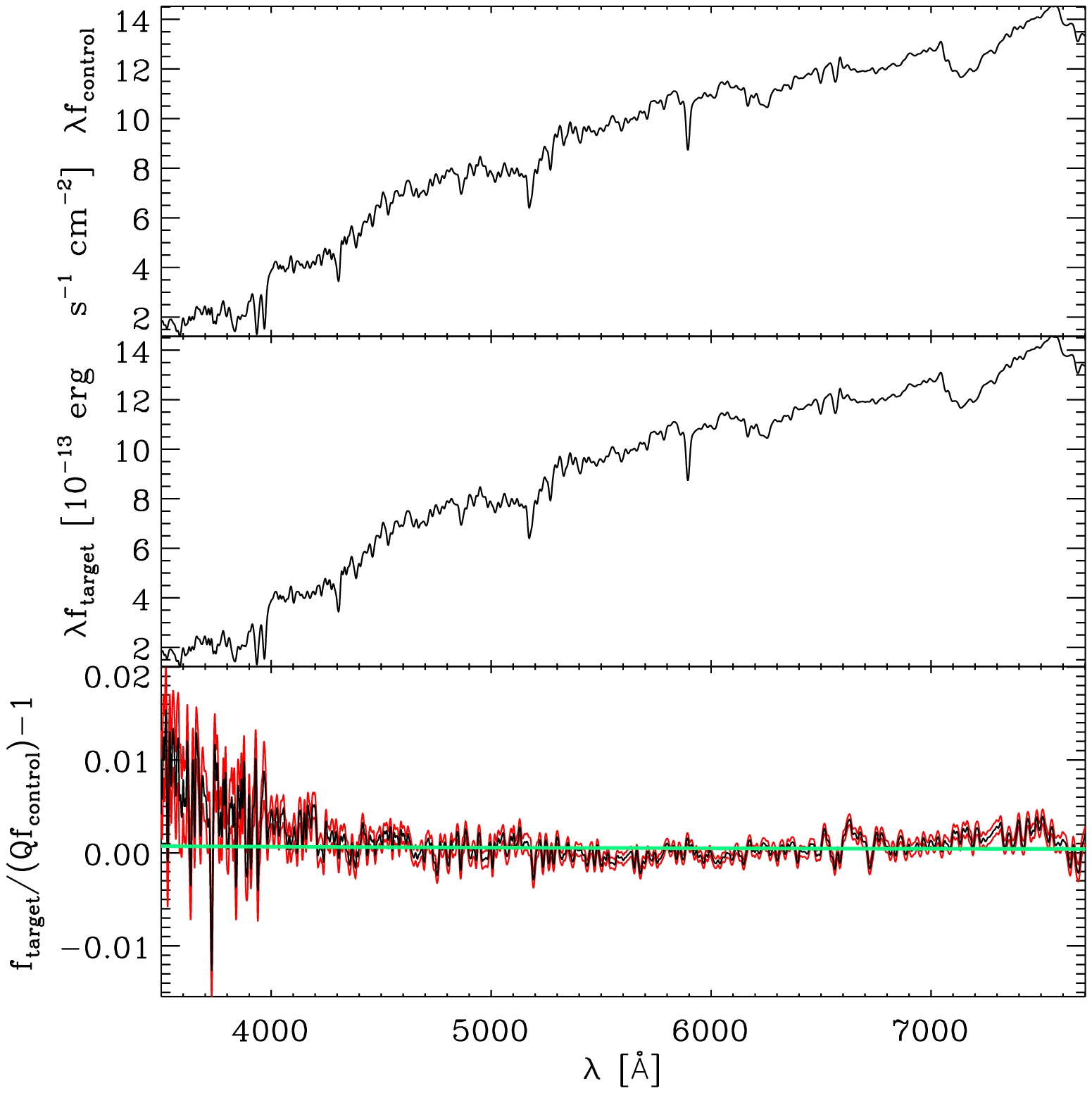}%
\caption{Results for the secondary samples: Each of these figures shows for the various secondary samples what was shown in Figure \ref{mainresult} for the primary sample. From left to right, top to bottom: $150 \leq \sigv \leq 200$; $250 \leq \sigv \leq 300$; \mingals\ = 5; \mingals\ = 20; \Rtarget\ = 0.25 Mpc; \Rtarget\ = 1 Mpc.}%
\label{sampleresults}
\end{figure*}

\begin{figure*}%
\centering
\includegraphics[height=0.3\textwidth]{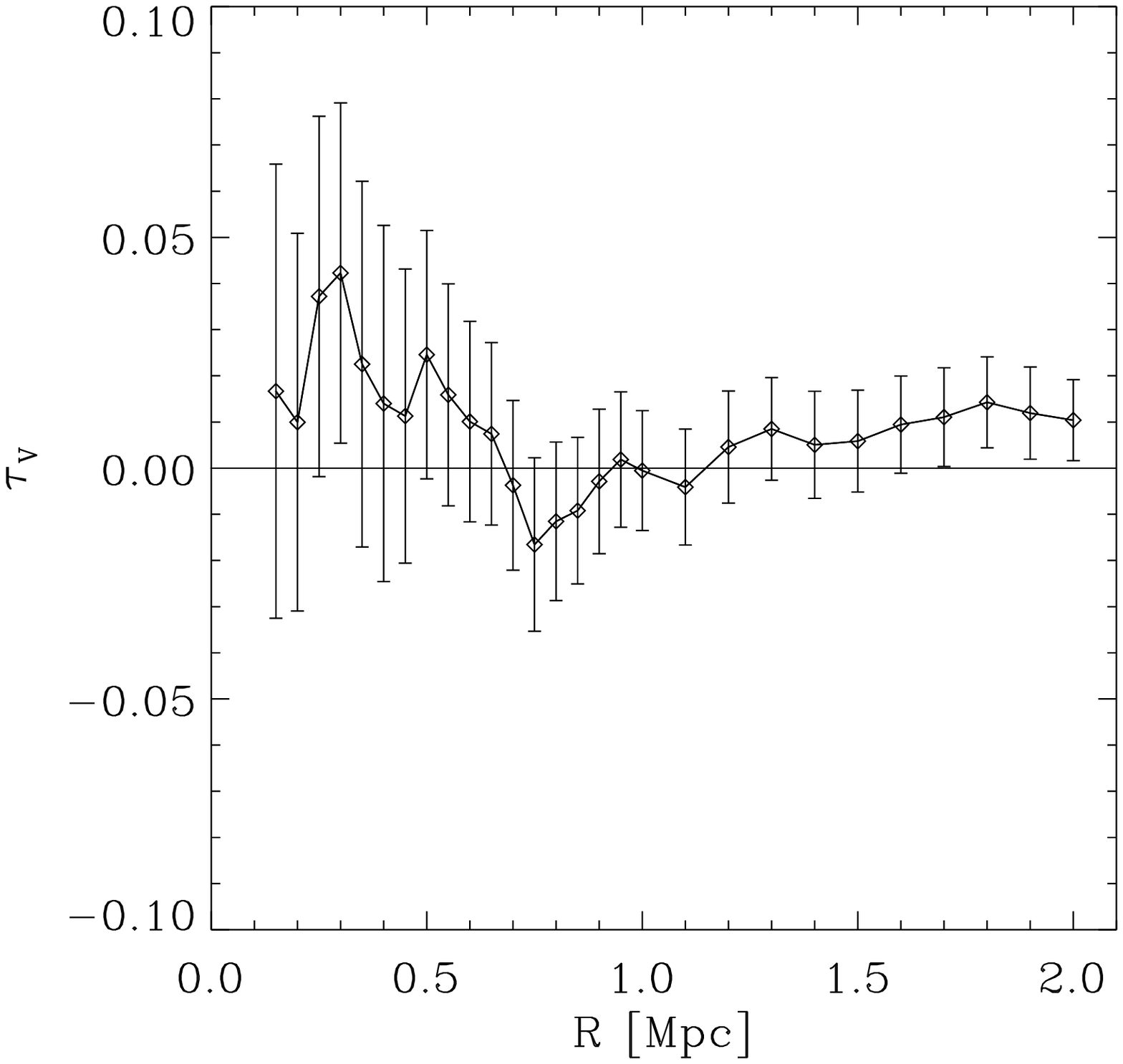}%
\includegraphics[height=0.3\textwidth]{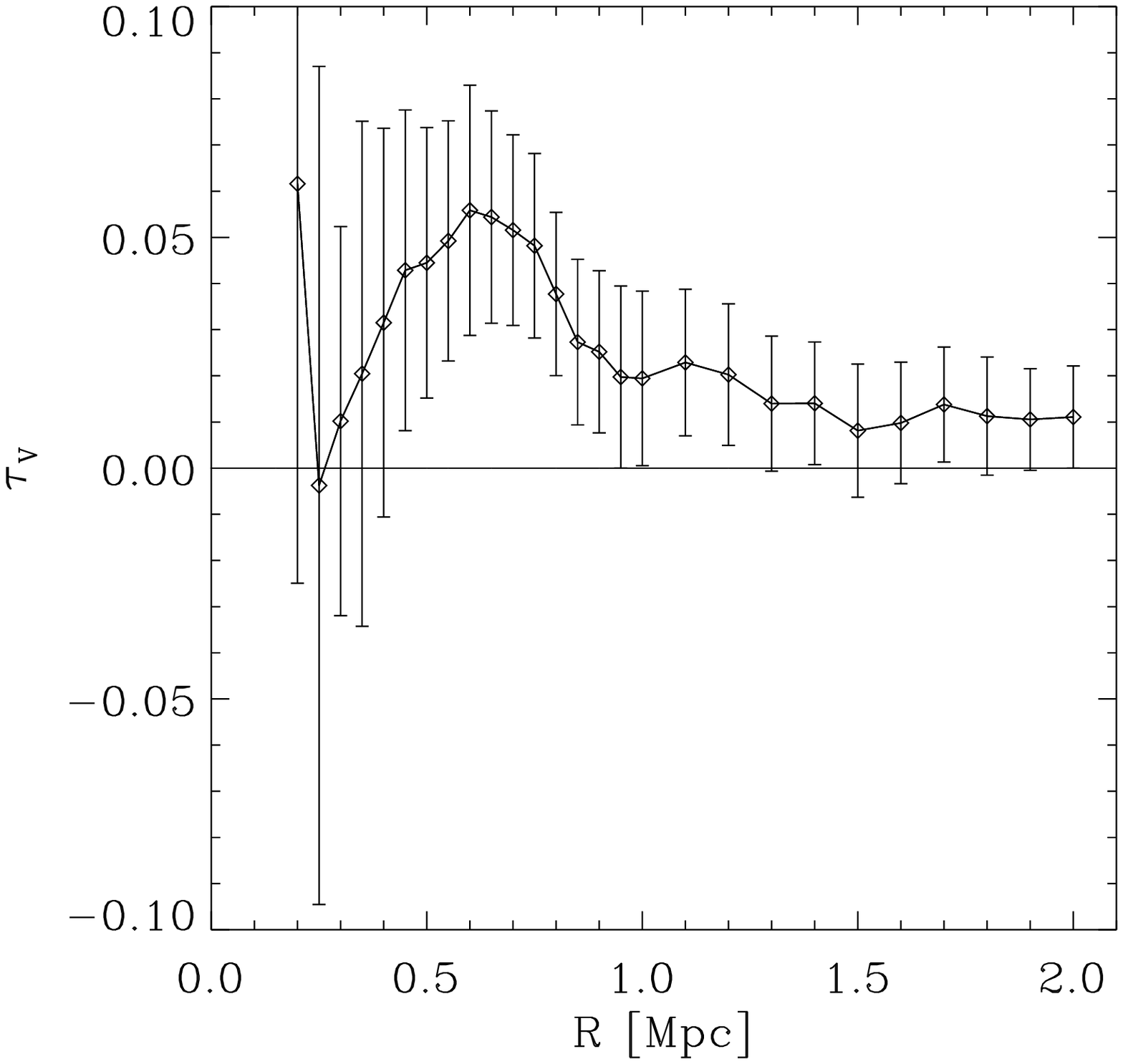}\\
\includegraphics[height=0.3\textwidth]{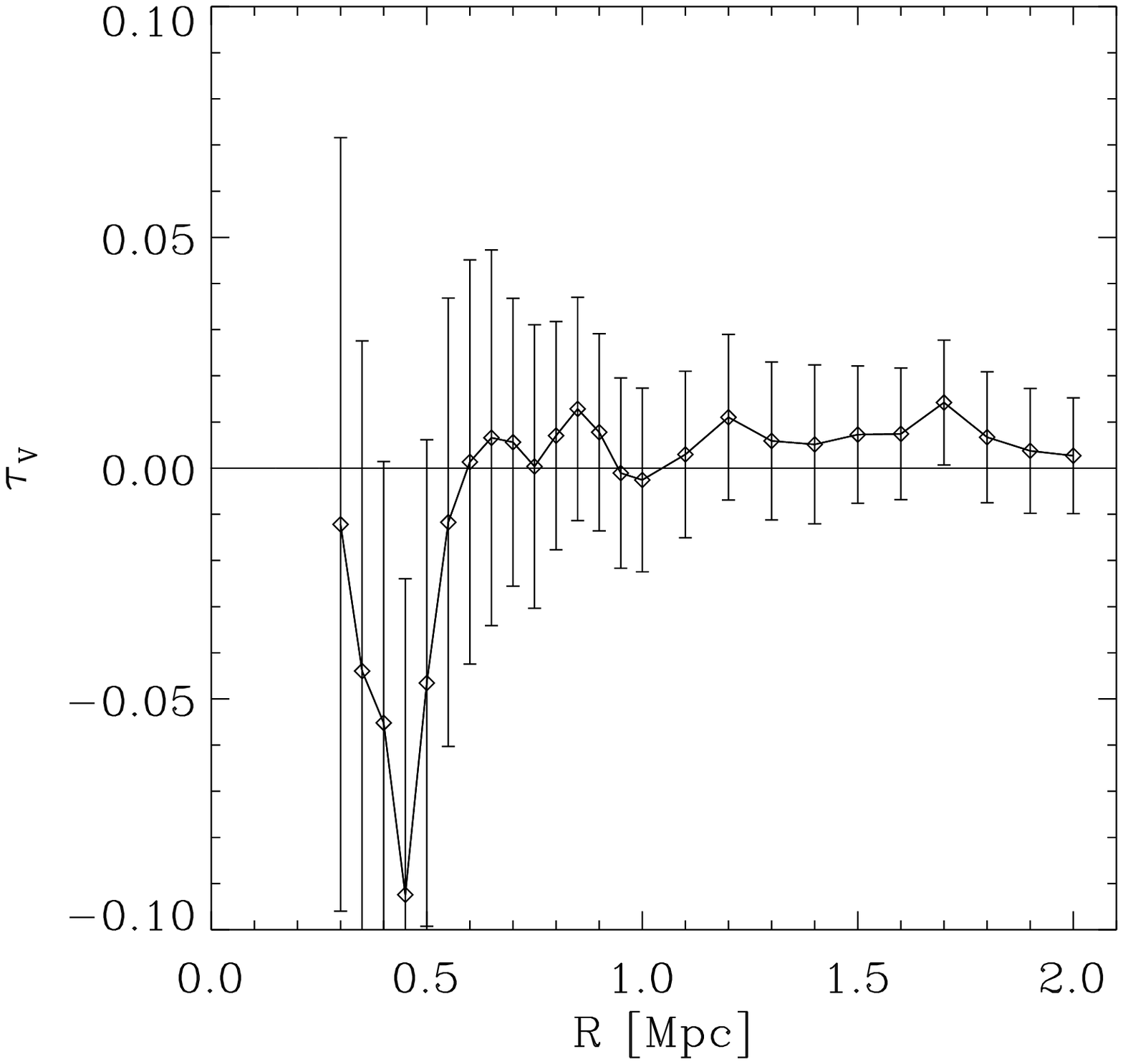}%
\includegraphics[height=0.3\textwidth]{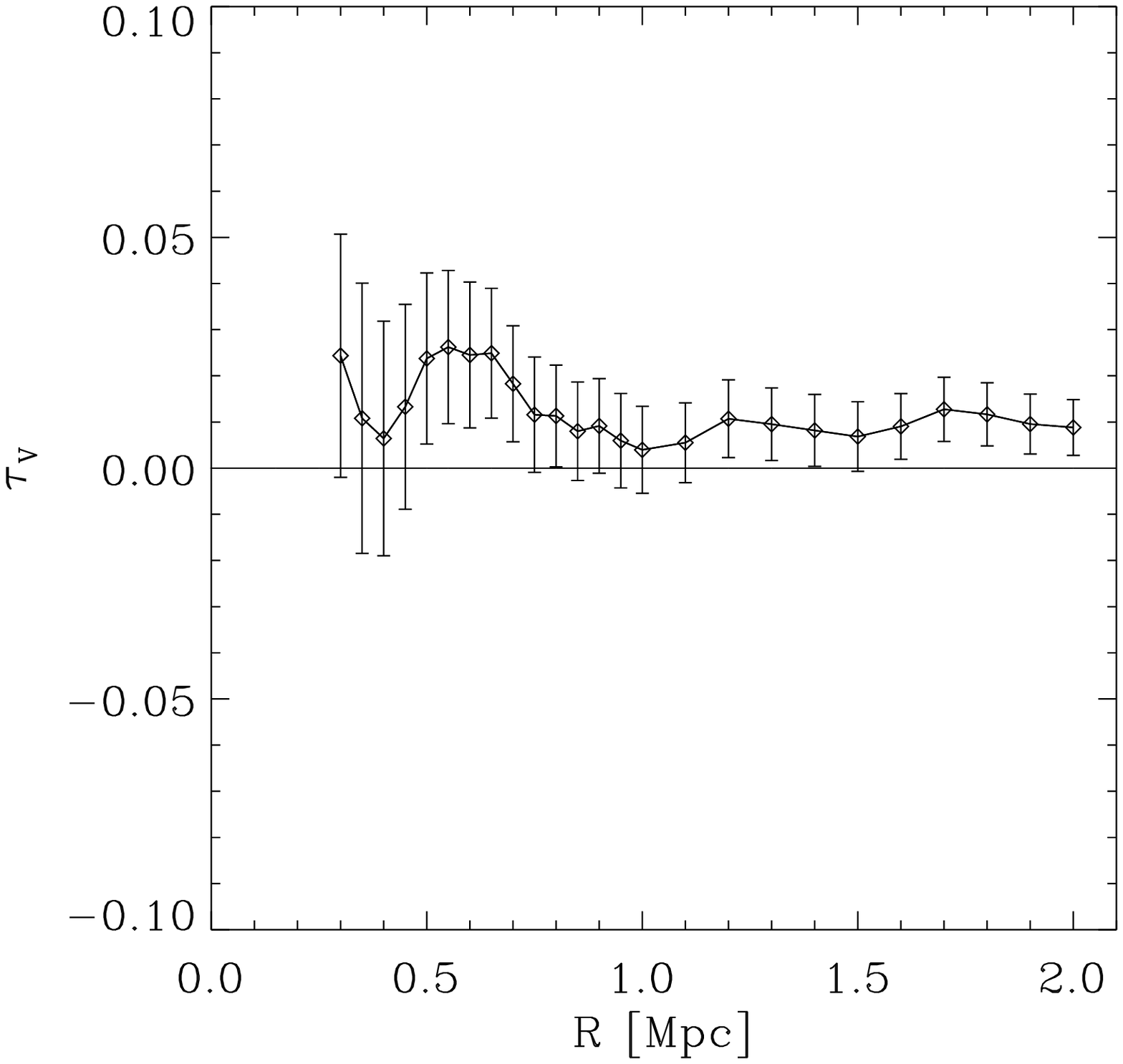}%
\caption{Radial dependence of the dust attenuation: radial dependence of the parameter \tauv\ in the dust law (eq. \ref{eqcomp}) for three velocity dispersion bins and their combined signal: top-left: $200 \leq \sigv \leq 250$; top-right: $150 \leq \sigv \leq 200$; bottom-left: $250 \leq \sigv \leq 300$; bottom-right: average of the three velocity dispersion bins. $R$ sets \Rtarget;  \mingals\ for each of these bins is 10, other parameters as for the primary sample. Note that the error bars in each plot are all covariant, since each datapoint is the cumulative signal from all galaxies inside \Rtarget.}%
\label{radialresults}
\end{figure*}

Figure \ref{sampleresults} gives the same analysis as Figure 
\ref{mainresult} for the secondary samples (see also Table 
\ref{tableresults}). Inspection of this figure and the results in the 
table shows that for smaller \target\ subsamples, the errors are 
significantly larger. 

We computed the radial dependence of \tauv\ for the 
various velocity dispersion bins by varying the value of \Rtarget\ for 
these samples, while keeping the value of \Rcontrol\ and other parameters 
fixed. The result is shown in Figure \ref{radialresults}. The plots shown are 
cumulative in the sense that each value of \Rtarget\ gives dust attenuation 
for dust within a volume of radius \Rtarget. The range between 0.15 and 1 Mpc 
was examined more carefully by increasing \Rtarget\ in steps of 0.05 Mpc, 
whereas between 1 and 2 Mpc, steps of 0.1 Mpc were used. For smaller values 
of \Rtarget\ the \target\ subsample consists of only a few galaxies, for which 
the \composite\ spectrum cannot be reliably obtained.

\section{Discussion}

The lower panel of Figure \ref{mainresult} shows a value of the
difference that is nearly flat over the whole wavelength range, and the
1-sigma error estimate on the value of \tauv\ in Table \ref{tableresults}
confirms that a null value is within the uncertainties. The other
samples confirm this result: Most of the samples give results that are
well within 1$\sigma$ of the null hypothesis, with a few of the samples
giving a formally negative value for \tauv, i.e. a negative
absorption. None of the values of \tauv\ are statistically significant
indicators of positive or negative absorption. Regarding the
negative absorptions, it must be remarked that the two most negative
values are obtained from relatively small \target\ subsample sizes,
about 30 spectra in the \target\ subsample (they both have the largest
error values as well), which could account for an estimate that is
significantly off. The other negative value is essentially zero and occurs
for the \Rtarget\ = 1 Mpc sample, which could 
simply indicate that, generically, at this radial distance there is no
dust in galaxy clusters (see below for a discussion of the radial
dependence).

The dust law used in equation (\ref{eqcomp}) is related to the extinction by
\begin{equation}
A(\lambda) = \tauv \times \frac{2.5}{\ln 10} \Big(\frac{\lambda}{5500 \mbox{\AA}} \Big)^{-0.7} \mbox{mag} \ ,
\end{equation}
which gives values of \EBV\ for the \tauv\ values obtained of the
order of 10$^{-3}$ mag. Our estimate of the typically error on
$\tauv$, $\sim0.03$, sets an upper bound on \EBV\ of $\sim\! 5 \times$
10$^{-3}$ mag for values of \Rtarget\ = 0.5 Mpc. Comparing these errors to 
what we found for our mock-samples in Section \ref{secmock}, we see that the 
obtained errors are consistent with being true measurement errors. The most 
positive value of \tauv\ is found for the $150 \leq \sigv \leq 200$ sample, 
which gives a reddening of \EBV\ = ($8 \pm 5$) $\times$ 10$^{-3}$ mag. The most
statistically significant value of \tauv\ is found for the \mingals = 5
 sample, with a value of \tauv\ that is 2$\sigma$
from the null result, corresponding to \EBV\ = ($5.0 \pm 2.5$) $\times$
10$^{-3}$ mag. These are still not very significant, but they are
remarkable in that they are obtained for samples that have both a large
\target\ and a large \control\ subsample.

The radial dependence plots are the most instructive of the resulting
plots as they might reveal the radial location of the dust content of
galaxy clusters. If there were dust in a certain radial distance
range, we would expect the \tauv\ vs. \Rtarget\ plot to be essentially
zero up to the dust range, after which a sharp increase would occur
over the range in which the dust occurs, followed by a gradual decline
for larger values of \Rtarget\ as more and more unattenuated spectra
are added to the \target\ subsample. For example, to confirm the
results of \cite{Chelouche:2007rm}, who found evidence of dust around
1 Mpc with no dust at smaller radii, we would expect to see a peak
around 1 Mpc.

In the range $< 1$ Mpc we do not find a consistent result in the three 
velocity dispersion bins we considered. The errors in this range are large 
because of the small number of galaxies in the \target\ subsamples of these 
bins ($\sim\!100$ for \Rtarget$\sim\! 0.5$ Mpc). The value of \tauv\ in the 
range $150 \leq \sigv \leq 250$ is positive; however, \tauv\ is in the 
negative range in the $250 \leq \sigv \leq 300$ bin. The amount of 
absorption rises in the interval between \Rtarget\ = 0.3 Mpc and 
\Rtarget\ = 0.7 Mpc for the $150 \leq \sigv \leq 200$ bin, but the opposite 
happens for the $200 \leq \sigv \leq 250$ bin. In the bottom-right panel of 
Figure \ref{radialresults}, we see that the combined result of the three 
bins is essentially flat within the error range.

The results are more consistent in the range between 1 and 2 Mpc. All of the 
bins show a small amount of dust absorption, but all the values 
contain the null value in their error range. The overall significance of the 
result is only slightly smaller than the significance of an individual point, 
because of strong correlations between different \Rtarget\ values (since this 
is a cumulative plot, the overlap between the different samples 
is significant). Combining the results in the 1-2 Mpc range, we find an 
average extinction \EBV\ = 0.002 mag with a significance of 1.5$\sigma$. At 
the 99\% confidence level, we conclude that \EBV\ $< 3 \times 10^{-3}$ mag.

A similar analysis for \Rtarget\ $\sim\!0.5$ Mpc gives an average extinction 
\EBV\ = 0.004 mag, with a significance of 1.2$\sigma$. Therefore, in this 
range we can derive a limit \EBV\ $< 8 \times 10^{-3}$ mag. Both of these 
upper bounds are more stringent than the ones previously found.

We can translate an upper bound on the extinction into an upper bound on the 
dust mass using \citep{Krugel03}

\begin{equation}
M_{\mathrm{dust}} = 1.5 \times 10^8 \frac{\EBV}{3 \times 10^{-3}
  \mathrm{ \ mag}}\Big( \frac{\mathcal{R}}{1 \mathrm{\ Mpc}} \Big)^2
M_\odot \ ,   
\end{equation}

\noindent which gives an upper bound $M_{\mathrm{dust}} \lesssim 10^8
M_\odot$ for $\sim\!$ Mpc-scales.

The bottom-right panel of Figure \ref{radialresults} summarizes our results: 
over the whole range we considered, \Rtarget\ $\sim\!0.15 - 2$ Mpc, we find a 
signal that is essentially flat and consistent with zero. Our upper limit on 
\EBV\ for distances between 1 and 2 Mpc from the center of a cluster is a 
few 10$^{-3}$  mag, which is consistent with the amount of dust extinction 
observed in $0.1 < z < 0.3$ clusters in \cite{Chelouche:2007rm}. Future work 
that could significantly increase the size of the \target\ subsample could 
lower the upper bound found here, or confirm the existence of dust on the 
outskirts of galaxy clusters.

\acknowledgements It is a pleasure to thank Mike Blanton for helpful
discussions and the anonymous referee for valuable comments.  This
project made use of the NASA Astrophysics Data System, and the
idlutils codebase maintained by David Schlegel and others.  Financial
support for this project was provided by the US National Aeronautics
and Space Administration (LTSA grant NAG5-11669, ADP grant
07-ADP07-0099, GALEX grant 06-GALEX06-0030, and Spitzer grant
G05-AR-50443).  During part of the period in which this research was
performed, DWH was a research fellow of the Alexander von Humboldt
Foundation of Germany.

Funding for the SDSS and SDSS-II has been provided by the Alfred P. Sloan 
Foundation, the Participating Institutions, the National Science Foundation, 
the U.S. Department of Energy, the National Aeronautics and Space 
Administration, the Japanese Monbukagakusho, the Max Planck Society, and the 
Higher Education Funding Council for England. The SDSS Web Site is 
\url{http://www.sdss.org/}.

The SDSS is managed by the Astrophysical Research Consortium for the 
Participating Institutions. The Participating Institutions are the American 
Museum of Natural History, Astrophysical Institute Potsdam, University of 
Basel, University of Cambridge, Case Western Reserve University, University 
of Chicago, Drexel University, Fermilab, the Institute for Advanced Study, 
the Japan Participation Group, Johns Hopkins University, the Joint Institute 
for Nuclear Astrophysics, the Kavli Institute for Particle Astrophysics and 
Cosmology, the Korean Scientist Group, the Chinese Academy of Sciences 
(LAMOST), Los Alamos National Laboratory, the Max-Planck-Institute for 
Astronomy (MPIA), the Max-Planck-Institute for Astrophysics (MPA), New 
Mexico State University, Ohio State University, University of Pittsburgh, 
University of Portsmouth, Princeton University, the United States Naval 
Observatory, and the University of Washington.

\end{document}